\documentclass[aps,prx,twocolumn,floatfix]{revtex4-2}
\usepackage{graphicx}
\usepackage{multirow}
\usepackage{color}
\usepackage{amsmath}
\usepackage{amsfonts}
\usepackage{amssymb}
\usepackage{graphicx}
\graphicspath{{Figures/}}
\usepackage{epstopdf}
\usepackage{empheq}
\usepackage{float}
\usepackage{cases}
\usepackage{mathtools}
\usepackage{dcolumn}
\usepackage{bbold}
\usepackage{xfrac}
\usepackage{bm}

\usepackage[dvipsnames]{xcolor}
\usepackage[colorlinks]{hyperref}
\hypersetup{
colorlinks=True,
linkbordercolor=White,
linkcolor=BrickRed,
citecolor=Blue,	 
}

\DeclarePairedDelimiter\bra{\langle}{\rvert}
\DeclarePairedDelimiter\ket{\lvert}{\rangle}

\newcommand{\gt}{\mathsf{g}}

\newcommand\YMN[1]{#1}

\begin{document}

\title{Hole spin manipulation in inhomogeneous and non-separable electric fields}

\author{Biel Martinez}
\thanks{These authors equally contributed to the work.}
\author{Jos\'e Carlos Abadillo-Uriel}
\thanks{These authors equally contributed to the work.}
\author{Esteban A. Rodr\'iguez-Mena}
\thanks{These authors equally contributed to the work.}
\author{Yann-Michel Niquet}
\email{yniquet@cea.fr}
\affiliation{Univ. Grenoble Alpes, CEA, IRIG-MEM-L\_Sim, Grenoble, France.}%

\date{\today}

\begin{abstract}
The usual models for electrical spin manipulation in semiconductor quantum dots assume that the confinement potential is separable in the three spatial dimensions and that the AC drive field is homogeneous. However, the electric field induced by the gates in quantum dot devices is not fully separable and displays significant inhomogeneities. Here, we address the electrical manipulation of hole spins in semiconductor heterostructures subject to inhomogeneous vertical electric fields and/or in-plane AC electric fields. We consider Ge quantum dots electrically confined in a Ge/GeSi quantum well as an illustration. We show that the lack of separability between the vertical and in-plane motions gives rise to an additional spin-orbit coupling mechanism (beyond the usual linear and cubic in momentum Rashba terms) that modulates the principal axes of the hole gyromagnetic $\gt$-matrix. This non-separability mechanism can be of the same order of magnitude as Rashba-type interactions, and enables spin manipulation when the magnetic field is applied in the plane of the heterostructure even if the dot is symmetric (disk-shaped). More generally, we show that Rabi oscillations in strongly patterned electric fields harness a variety of $\gt$-factor modulations. We discuss the implications for the design, modeling and understanding of hole spin qubit devices. 
\end{abstract}

\maketitle

\section{Introduction}

Hole spin qubits in semiconductor quantum dots afford the unique advantage of an efficient electrical control \cite{Burkard22}. This control is enabled by the strong spin-orbit interaction (SOI) in the valence band of semiconductors, which couples the spin to the real-space motion of the hole in the applied electric fields \cite{Winkler03,Kloeffel11,Kloeffel18}. Rabi (spin rotation) frequencies in the tens of MHz range are thus routinely achieved in hole spin qubit devices \cite{Maurand16,Crippa18,Watzinger18,Hendrickx20b,Camenzind22,Froning21,wang2022ultrafast}. This electrical spin susceptibility, however, comes at the expense of a stronger sensitivity to charge noise and disorder \cite{Martinez2022}. Yet considerable progress has been made recently, with the theoretical and experimental demonstration of operational sweet spots where the hole spins decouple from longitudinal (dephasing) noise and, therefore, show long coherence times while remaining electrically addressable \cite{Wang21,Bosco21,Piot22,michal2022tunable}. The versatile interactions between hole spins and electric fields also hold promises for strong spin-photon coupling, opening new opportunities for circuit quantum electro-dynamics applications such as long-range spin-spin interactions \cite{Kloeffel13,michal2022tunable,Bosco22,yu2022strong}.

The early demonstrations of hole spin qubits in Si/SiO$_2$ devices have been recently outmatched by epitaxial Ge/GeSi heterostructures \cite{Scappucci20,Hendrickx20b,Hendrickx20,Hendrickx21}. In such heterostructures, a thin Ge quantum well hosting the holes is buried $20-50$\,nm below the surface on which the gates shaping and controlling the dots are deposited. This mitigates the impact of electrical and charge noise from the gate stack. Moreover, the holes are lighter in Ge than in Si, hence the characteristic confinement lengths are larger, which relaxes the constraints on dot size and gate pitch. A four spin qubits processor in a Ge/GeSi heterostructure has thus been demonstrated recently \cite{Hendrickx21}, and charge control has been achieved in a sixteen dots array \cite{Borsoi22}.

Such qubits can be manipulated by ``shaking'' the dot as a whole with an in-plane, time-dependent (AC) electric field resonant with the Zeeman spin splitting in a finite magnetic field $\mathbf{B}$. The SOI experienced by the hole indeed translates into an effective time-dependent magnetic field in the frame of the moving dot, which drives spin rotations \cite{Rashba03,Golovach06}. This SOI involves different heavy-hole (HH)/light-hole (LH) mixing terms depending on the symmetries of the dot. It is usually discussed with respect to paradigmatic situations where the motion of the hole in the $xy$ plane of the heterostructure is separable from the motion along the growth axis $z$ \YMN{ -- namely, the potential can be split as $V(x,y,z)\equiv V_\parallel(x,y)+V_\perp(z)$}. In highly symmetric (disk-shaped) quantum dots, the vertical electric field gives rise to a Rashba SOI that is cubic in the in-plane momentum components $p_x$ and $p_y$ \cite{Marcellina17,Terrazos21,Wang21}. However, this interaction does only harness the small anisotropy of the valence band of germanium. Structural asymmetries brought by the Si/Ge interfaces \cite{Xiong21,Liu22} or enforced by squeezing the dots laterally \cite{Michal21,Bosco21b} result in a Rashba SOI that is linear in $p_x$, $p_y$ and can easily outweigh the cubic term. The deformations of the moving dot may also modulate the gyromagnetic $\gt$-factors of the hole and make a so-called $\gt$-tensor modulation resonance ($\gt$-TMR) contribution to the Rabi frequency \cite{Martinez2022,Kato03,ares2013sige,Crippa18}.

In the experiment reported in Ref. \cite{Hendrickx21}, the static magnetic field is applied in-plane in order to minimize dephasing noise due to the hyperfine interactions with Ge isotopes carrying nuclear spins \cite{fischer2008spin,testelin2009hole,Bosco21c}. However, the cubic Rashba SOI does not allow for spin manipulation in this setup as the resulting Rabi frequency is proportional to the vertical magnetic field $B_z$ \cite{Marcellina17,Terrazos21}. Here, we show that the Rabi frequency of circular and mildly (but realistically) squeezed dots can indeed be dominated by a sharp in-plane feature. The latter results, in particular, from the non-separability (NS) of the in-plane and out-of-plane motions of the hole \YMN{[$V(x,y,z)\not\equiv V_\parallel(x,y)+V_\perp(z)$]}. This $\gt$-TMR-like contribution is enhanced in Ge/GeSi heterostructures by the large depth of the well that promotes an electrostatic tip effect from the gates and has been overlooked up to now. It is, however, not material-specific and shall be ubiquitous in a large variety of devices. We show, more generally, that Rabi oscillations in the highly patterned electric fields encountered in such devices harness a variety of $\gt$-TMR mechanisms that depend on the layout of the gates used to drive the dot. We discuss the practical consequences for the design, modeling and understanding of spin qubit devices in planar heterostructures.

\begin{figure}
    \centering
    \includegraphics[width=.85\columnwidth]{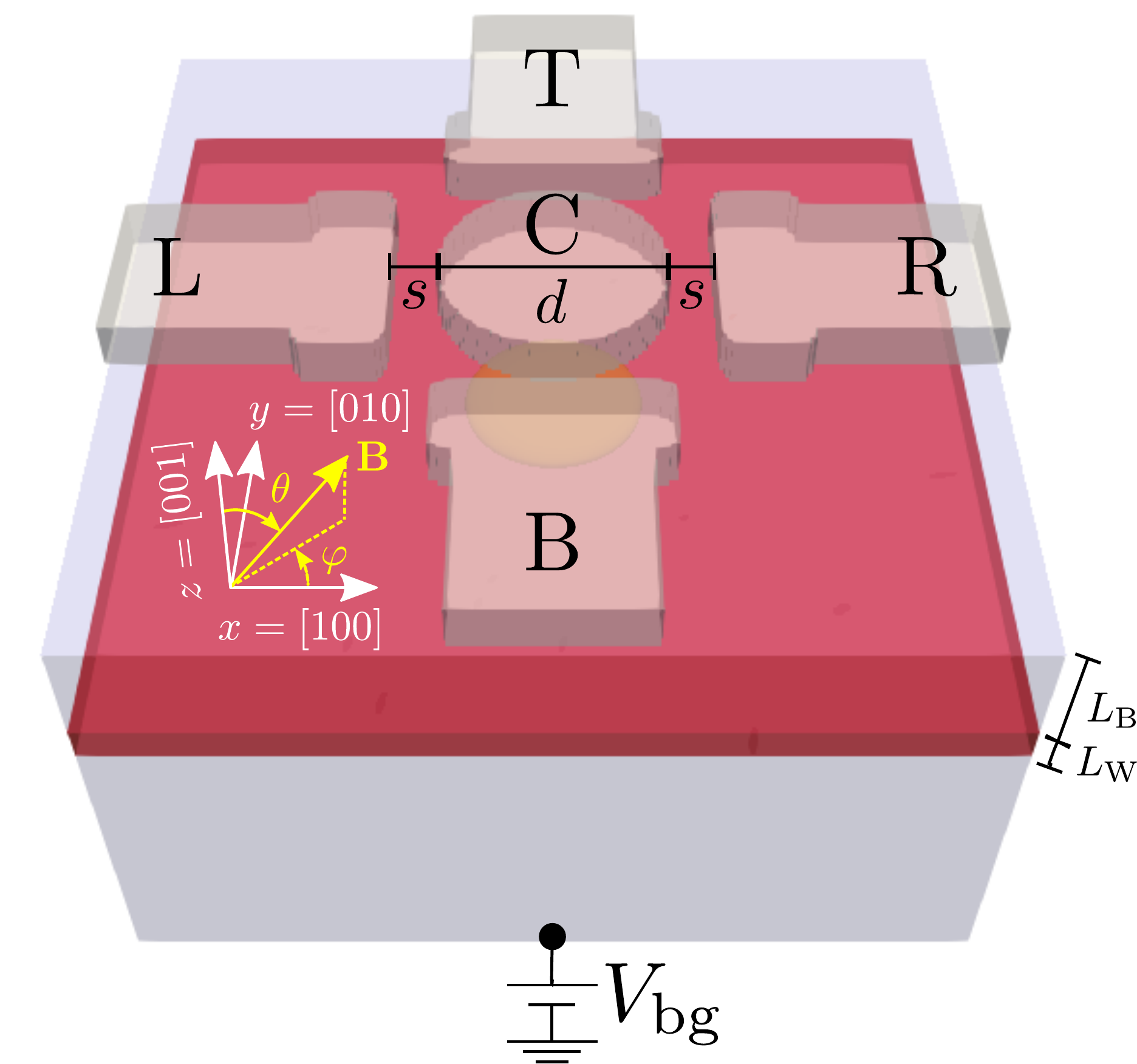}
    \caption{The hole spin qubit device. The Ge quantum well (red) is embedded between  Ge$_{0.8}$Si$_{0.2}$ barriers (light blue). The difference of potential between the central front gate (C) and the grounded side gates (L/R/T/B) shapes a hole quantum dot in this well. Unless otherwise specified, the Ge well is $L_\mathrm{W}=16$\,nm thick, the upper barrier $L_\mathrm{B}=50$\,nm thick, the diameter of the C gate is $d=100$ nm and the gap with the side gates is $s=20$ nm. All gates are embedded in Al$_2$O$_3$, and are insulated from the heterostructure by 5\,nm of this material. The substrate below the 150 nm thick lower barrier acts as an effective back gate, used to tune independently the depth of the quantum dot and the vertical electric field. We assume, as in Ref.~\cite{Sammak19}, that the Ge$_{0.8}$Si$_{0.2}$ barriers are not fully relaxed, and experience residual in-plane strain $\varepsilon_{xx}=\varepsilon_{yy}=\varepsilon_\parallel=0.26\%$ and out-of-plane strain $\varepsilon_{zz}=\varepsilon_\perp=-0.19\%$. Consequently, the strains in the Ge well are $\varepsilon_\parallel=-0.63\%$ and $\varepsilon_\perp=+0.47\%$. The yellow contour is the isodensity surface that encloses 90\% of the ground-state hole charge at $V_\mathrm{C}=-40$\,mV and $V_\mathrm{bg}=0$\,V.}
    \label{fig:device}
\end{figure}

\section{Rabi oscillations in a non-separable potential}

We highlight the relevance of non-separability on the device of Fig.~\ref{fig:device}. The latter can be viewed as the elementary tile of a 2D array of hole spin qubits in a strained Ge/GeSi heterostructure similar to Ref.~\cite{Hendrickx21}. A single hole is confined under the central gate C by the difference of potential with the grounded side gates. The spin of the hole is electrically manipulated by opposite AC modulations $\delta V_\mathrm{L}=-\delta V_\mathrm{R}=(V_\mathrm{ac}/2)\cos\omega_\mathrm{L}t$ on the L and R gates. These modulations aim to ``shake'' the dot as a whole along the $x$ axis as in the usual arrangements proposed to leverage Rashba SOI \cite{Marcellina17,Terrazos21,Wang21,Bosco21b}. We compute the static and AC potentials by solving Poisson's equation with a finite volumes method, then the ground-state wave functions with a finite differences, four bands Luttinger-Kohn (LK) Hamiltonian \cite{Luttinger56,KP09} accounting for the effects of the magnetic field on the orbital and spin degrees of freedom (see Appendix \ref{app:hamiltonian}). Finally, we extract the spin Rabi frequency from the dependence of the gyromagnetic $\gt$-matrix of the hole on the gate voltages \cite{Venitucci18}.

\begin{figure}
    \centering
    \includegraphics[width=1\columnwidth]{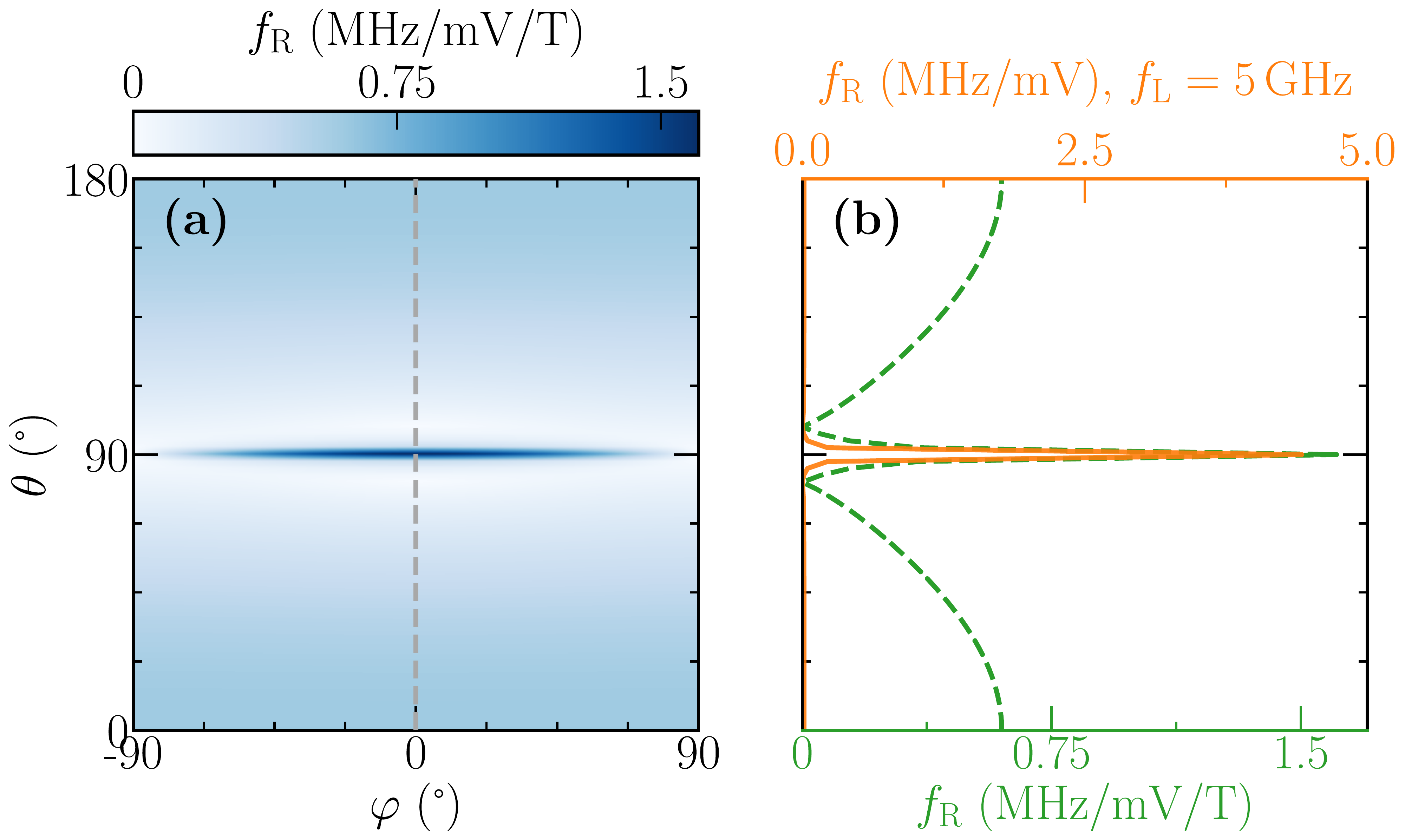}
    \caption{(a) Map of the Rabi frequency $f_\mathrm{R}$ as a function of the magnetic field angles $\theta$ and $\varphi$ defined in Fig.~\ref{fig:device}, for opposite drives $\delta V_\mathrm{L}=-\delta V_\mathrm{R}=(V_\mathrm{ac}/2)\cos\omega_\mathrm{L}t$ ($V_\mathrm{C}=-40$\,mV, $V_\mathrm{bg}=0$\,V, $V_\mathrm{ac}=1$\,mV and $B=1$\,T). (b) Cut along the dashed gray line in (a), at constant magnetic field $B=1$\,T (green), and at constant Larmor frequency $f_\mathrm{L}=\omega_\mathrm{L}/(2\pi)=5$\,GHz (orange).}
    \label{fig:Angularmap}
\end{figure}

In such a device, the ground-state has a strong heavy-hole character owing to the vertical confinement and to the compressive lattice-mismatch strains in the well \cite{Sammak19}. This is evidenced by highly anisotropic principal $\gt$-factors $\gt_z\approx 13.25$ and $\YMN{|\gt_x|=|\gt_y|=\gt_\parallel}$ ranging from $0.05$ to $0.15$. The in-plane $\gt$-factors are actually strongly dependent on the dot radius $r_\parallel=\sqrt{\langle x^2\rangle+\langle y^2\rangle}$ (see Appendix \ref{app:hamiltonian}), being minimal in the smallest dots (large $V_\mathrm{C}\ll0$, $r_\parallel\simeq15$ nm) and maximal in the largest ones ($V_\mathrm{C}\lesssim 0$, $r_\parallel\simeq30$ nm). This is consistent with the trends shown in Ref.~\cite{Wang22}, given that the dots are slightly smaller here. The Rabi frequency computed at $V_\mathrm{C}=-40$\,mV and $V_\mathrm{bg}=0$\,V is plotted as a function of the orientation of the magnetic field $\mathbf{B}$ in Fig.~\ref{fig:Angularmap}. The Rabi frequencies being proportional to the magnetic field strength $B$ and AC drive voltage $V_\mathrm{ac}$, they are normalized to $B=1$ T and $V_\mathrm{ac}=1$\,mV. Note that the \YMN{AC electric field in the dot, $E_{\mathrm{ac},x}\approx 1.7$\,$\mu$V/nm at $V_\mathrm{ac}=1$\,mV,} is much smaller than in typical Si/SiO$_2$ devices \cite{Venitucci18,Michal21} owing to the pitch of the gates and the depth of the well. Strikingly, this map shows the $\propto B_z\propto\sin\theta$ background expected for cubic Rashba SOI \cite{Terrazos21,Wang21}, yet outweighed by an extra, prominent in-plane feature ($\theta=90^\circ$). The Rabi frequency is actually maximal when the magnetic field is along $x$, where it reaches $f_\mathrm{R}=1.5$ MHz/mV/T. The sharp in-plane peak stands out even more if the Rabi frequencies are plotted at constant Larmor frequency $f_\mathrm{L}=\omega_\mathrm{L}/2\pi=\mu_B\sqrt{\gt_x^2B_x^2+\gt_y^2B_y^2+\gt_z^2B_z^2}/h$ (with $\mu_B$ the Bohr magneton), given the large $\gt_z/\YMN{|}\gt_{x,y}\YMN{|}$ ratio. 

This feature results from the coupling between the in-plane and out-of-plane motions of the hole in the strongly patterned electric field of the gates. To evidence this, we start from the usual paradigm where the \YMN{confinement and AC potentials are} separable \footnote{\YMN{An example of separable confinement is an in-plane harmonic potential $V_\parallel(x,y)=m_\parallel\omega_\parallel^2(x^2+y^2)/2$ and a vertical electric field $V_\perp(z)=-eE_zz$ with hard-wall boundary conditions $|z|\le L_\mathrm{W}/2$ ($m_\parallel$ being the in-plane mass of the holes). An example of separable AC potential is a homogeneous AC electric field oriented along $\mathbf{x}$ (or whatever direction).}}. In the absence of HH/LH mixing, the vertical confinement gives rise to pure heavy ($\ket{n_z,\pm\tfrac{3}{2}}$) and light ($\ket{n_z,\pm\tfrac{1}{2}}$) hole subbands. These are mixed by the so-called $R$ and $S$ terms of the LK Hamiltonian \cite{Luttinger56}, and by the off-diagonal elements of the hole Zeeman Hamiltonian $H_\mathrm{Z}=2\mu_B(\kappa\mathbf{B}\cdot\mathbf{J}+q\mathbf{B}\cdot\mathbf{J}^3)$, where $\mathbf{J}$ is the spin $\tfrac{3}{2}$ operator, $\mathbf{J}^3\equiv(J_x^3,J_y^3,J_z^3)$, and $\kappa$, $q$ are the isotropic and cubic Zeeman parameters (see Appendix \ref{app:hamiltonian}) \footnote{We use $\gamma_1=13.38$, $\gamma_2=4.24$, $\gamma_3=5.69$, $\kappa=3.41$, $q=0.06$, $b_v=-2.16$\,eV and $\nu=0.75$ in Ge; and $\gamma_1=11.56$, $\gamma_2=3.46$, $\gamma_3=4.84$, $\kappa=2.64$, $q=0.05$, $b_v=-2.19$\,eV and $\nu=0.76$ in Ge$_{0.8}$Si$_{0.2}$ \cite{Winkler03,Fischetti96}.}. $R\propto p_xp_y, p_x^2-p_y^2$ couples the in-plane motion of the heavy and light holes, while $S\propto\gamma_3(p_xp_z-ip_yp_z)$ couples their in-plane and out-of-plane motions (here $\mathbf{p}$ is the hole momentum and the $\gamma$'s are the Luttinger parameters). The effective Hamiltonian ${\cal H}$ for the in-plane motion in the lowest HH subband $n_z=0$ can then be obtained by a Schrieffer-Wolff transformation integrating out the LH subbands. Discarding the coupling with the farther-lying $n_z>0$ LH subbands for simplicity, we find
\begin{equation}
{\cal H}_{hh^\prime}\approx-\frac{1}{\Delta_\mathrm{LH}}\sum_{l} \bra{0,h}H_\mathrm{c}\ket{0,l}\bra{0,l}H_\mathrm{c}^\prime\ket{0,h^\prime}\,.
\end{equation}
Here $h,h^\prime=\pm\tfrac{3}{2}$, $l=\pm\tfrac{1}{2}$, $\Delta_\mathrm{LH}$ is the splitting between the ground HH and LH subbands, and $H_\mathrm{c},\,H_\mathrm{c}^\prime\in\{R,\,S,\,H_\mathrm{Z}\}$ are the HH/LH mixing terms. Setting $H_\mathrm{c}=S$, $H_\mathrm{c}^\prime=R$ (or vice-versa) yields the cubic Rashba SOI in symmetric dots \cite{Marcellina17}, and the linear Rashba SOI in squeezed dots \cite{Bosco21b}. Setting otherwise $H_\mathrm{c}=S$, $H_\mathrm{c}^\prime=H_\mathrm{Z}$ (or vice-versa) couples $h=+\tfrac{3}{2}$ to $h^\prime=+\tfrac{3}{2}$ through virtual $l=+\tfrac{1}{2}$ excitations, and $h=-\tfrac{3}{2}$ to $h^\prime=-\tfrac{3}{2}$ through $l=-\tfrac{1}{2}$. This gives rise to an effective interaction ${\cal H}_{hh'}\approx \beta(\kappa\gamma_3/\Delta_\mathrm{LH})(B_x p_y-B_y p_x)\delta_{hh'}$, where $\beta$ depends on the vertical confinement potential. This interaction has no action on the HH spin and can not, therefore, lead to Rabi oscillations.  

\begin{figure}
    \centering
    \includegraphics[width=1\columnwidth]{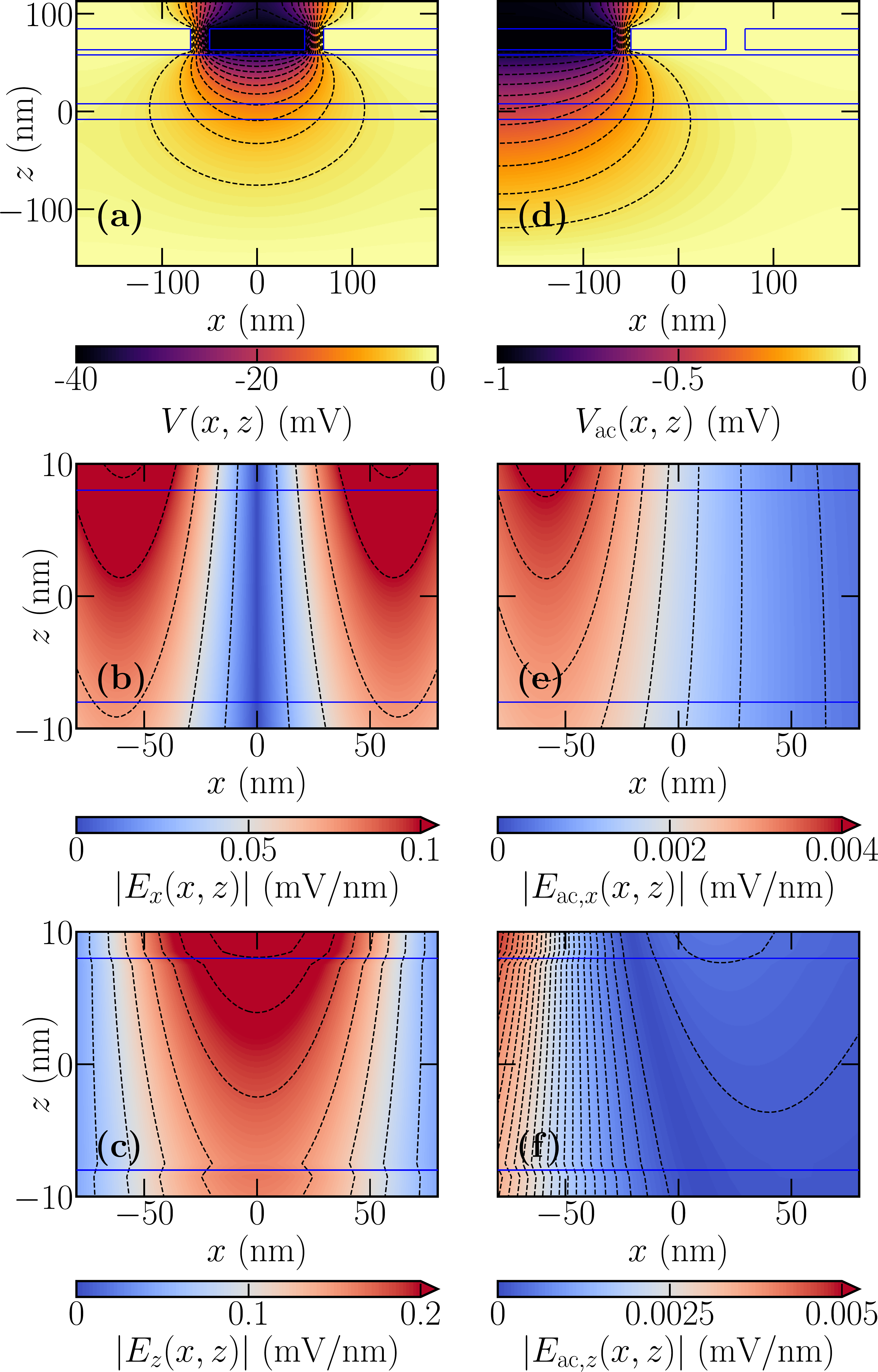}
    \caption{(a) Map of the electrical confinement potential $V(x,z)$ in the $xz$ symmetry plane at $y=0$, computed at $V_\mathrm{C}=-40$\,mV and $V_\mathrm{bg}=0$\,V. (b, c) Components of the electric field $E_x=-\partial V(x,z)/\partial x$ and $E_z=-\partial V(x,z)/\partial z$ (close up on the dot area). (d) Map of the AC potential $V_\mathrm{ac}(x,z)$ in the $xz$ plane at $y=0$, computed for a $V_\mathrm{L}=-1$\,mV pulse on the L gate. (e, f) Components of the AC electric field $E_{\mathrm{ac},x}=-\partial V_\mathrm{ac}(x,z)/\partial x$ and $E_{\mathrm{ac},z}=-\partial V_\mathrm{ac}(x,z)/\partial z$ (close up on the dot area). The blue lines delineate the different materials. The origin of coordinates is at the middle of the well along the C gate symmetry axis.}
    \label{fig:potlandscape}
\end{figure}

The situation is, however, different when the in-plane and vertical \YMN{confinement (and/or AC potentials)} are not separable. In that case, the dot deforms when shaken along $x$ or $y$ owing to the coupling between the in-plane and out-of-plane motions. \YMN{This gives rise to AC modulations of the HH/LH mixings by $S$, hence to AC modulations of the Zeeman interaction by $H_\mathrm{Z}$}, and to a $\propto B$ contribution to the Rabi frequency. A Schrieffer-Wolff transformation including the AC drive $\delta V(t)=\delta V_\mathrm{L}(t)-\delta V_\mathrm{R}(t)$ along with the $S$ and $H_\mathrm{Z}$ terms yields the minimal Hamiltonian that accounts for this effect in the ground-state heavy-hole doublet (see Appendices \ref{app:theory} and \ref{app:gTMR}):
\begin{equation}
{\cal H}=\frac{1}{2}\mu_B\boldsymbol{\sigma}\cdot\gt\mathbf{B}+\frac{1}{2}\mu_B \delta V(t)(\lambda_x B_x+\lambda_y B_y)\sigma_z\,,
\end{equation}
where $\boldsymbol{\sigma}$ is the vector of Pauli matrices, $\gt\approx\mathrm{diag}(\YMN{\gt_x=\gt_\parallel,\,\gt_y=-\gt_\parallel},\,\gt_z)$ is the gyromagnetic $\gt$-matrix of the dot, and the coupling constants $\lambda_{x,y}\propto\gamma_3\kappa/\Delta_\mathrm{LH}$ depend on the static and AC potentials. The $\propto \delta V(t)B_{x,y}\sigma_z$ form of the drive Hamiltonian is generic (irrespective of these potentials) because $S$ and the in-plane magnetic field in $H_\mathrm{Z}$ only connect $h=h^\prime$ states through virtual LH excitations; yet, as hinted above, the coupling constants $\lambda_x$ and $\lambda_y$ are zero if the \YMN{confinement and AC potentials} are both separable. \YMN{Note that $\gt_y=-\gt_x$ is actually negative in our $\{\ket{\tfrac{3}{2}},\,\ket{-\tfrac{3}{2}}\}$ basis set (see Appendix \ref{app:hamiltonian})}.

\begin{figure}
    \centering
    \includegraphics[width=1\columnwidth]{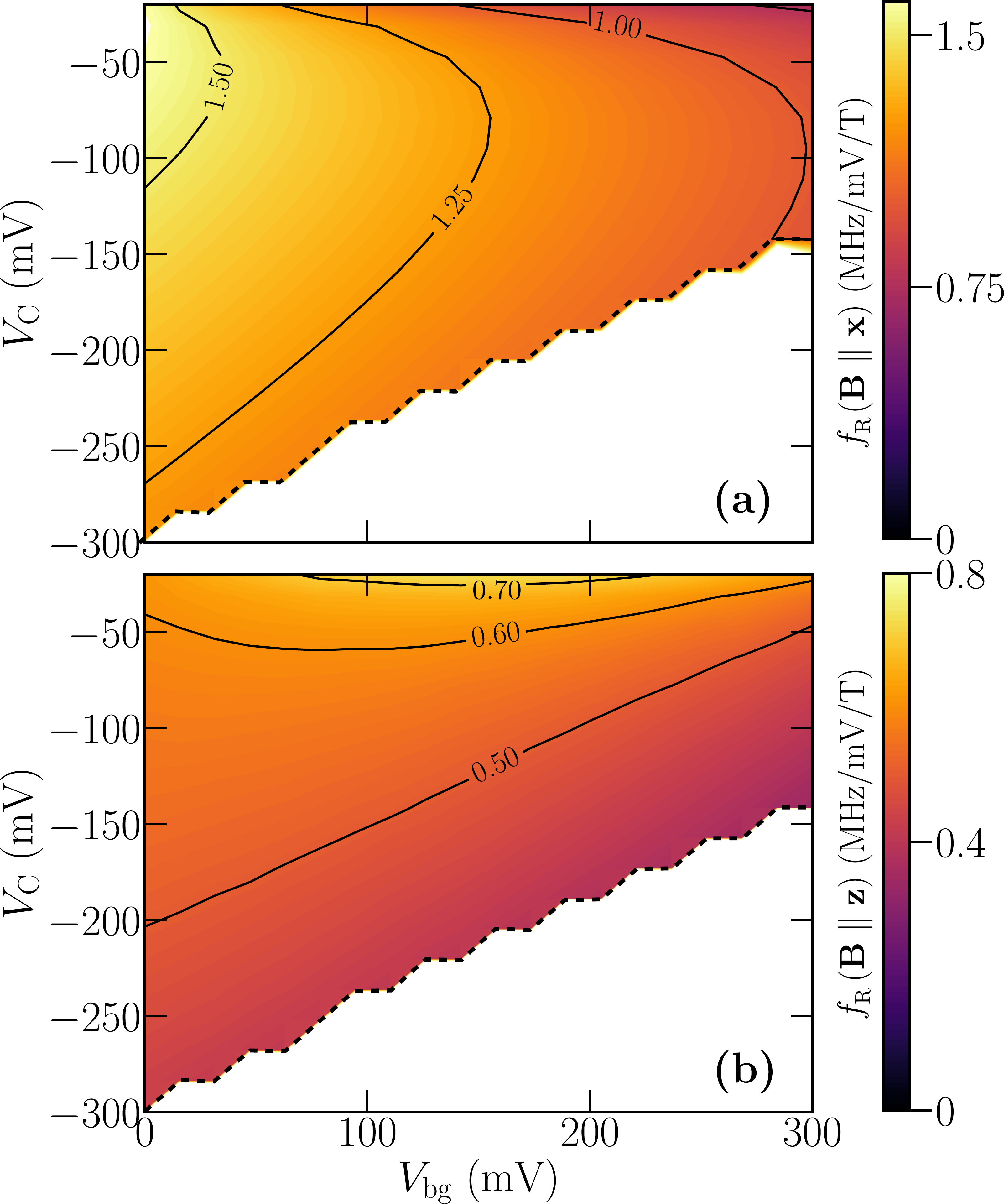}
    \caption{(a) Map of the Rabi frequency $f_\mathrm{R}(\mathbf{B}\parallel\mathbf{x})$ at constant magnetic field $B=1$\,T as a function of $V_\mathrm{C}$ and $V_\mathrm{bg}$ (opposite drives on the L and R gates). (b) Same for $f_\mathrm{R}(\mathbf{B}\parallel\mathbf{z})$. Note the different scales in (a) and (b). The hole is pulled by the electric field from the well to the top SiGe/Al$_2$O$_3$ interface in the white areas.}
    \label{fig:Vgmap}
\end{figure}

The $\propto\delta V(t)$ term can be cast as ${\cal H}_\mathrm{ac}(t)=\tfrac{1}{2}V_\mathrm{ac}\cos(\omega_\mathrm{L}t)\mu_B \boldsymbol{\sigma}\cdot\gt^\prime\mathbf{B}$, where $\gt^\prime=\partial\gt/\partial(V_\mathrm{L}-V_\mathrm{R})$ with $\gt_{zx}^\prime=\lambda_x$ and $\gt_{zy}^\prime=\lambda_y$; this $\gt^\prime$ matrix gives rise to Rabi oscillations with frequency \cite{Venitucci18}:
\begin{subequations}
\begin{align}
f_\mathrm{R}&=\frac{\mu_BBV_\mathrm{ac}}{2h\gt^*}\left|\gt\mathbf{b}\times\gt^\prime\mathbf{b}\right| \label{eq:fRg} \\
&=\frac{\mu_BBV_\mathrm{ac}}{2h\gt^*}\left|\gt_{zx}^\prime b_x+\gt_{zy}^\prime b_y\right|\sqrt{\gt_x^2b_x^2+\gt_y^2b_y^2}\,,
\label{eq:gTMR}
\end{align}
\end{subequations}
where $\gt^*=\sqrt{\gt_x^2b_x^2+\gt_y^2b_y^2+\gt_z^2b_z^2}$ is the effective $\gt$-factor of the dot and $\mathbf{b}=(b_x,b_y,b_z)$ is the unit vector along $\mathbf{B}$. The Rabi frequency is, therefore, strongly peaked for in-plane magnetic fields owing, again, to the large $\gt_z/\YMN{|}\gt_{x,y}\YMN{|}$ ratio. When the Rabi oscillations are driven with opposite modulations on the L and R gates of Fig.~\ref{fig:device}, $\gt_{zy}^\prime=0$ so that $f_\mathrm{R}$ is maximal when $\mathbf{B}\parallel\mathbf{x}$, and zero when $\mathbf{B}\parallel\mathbf{y}$. At variance with conventional $g$-TMR mechanisms also involving an interplay with the Zeeman and LK Hamiltonians \cite{Venitucci19,Michal21}, the drive does not directly modulate the principal $\gt$-factors $\gt_x$, $\gt_y$ and $\gt_z$, but the principal axes of the $\gt$-matrix.

\begin{figure}
    \centering
    \includegraphics[width=1\columnwidth]{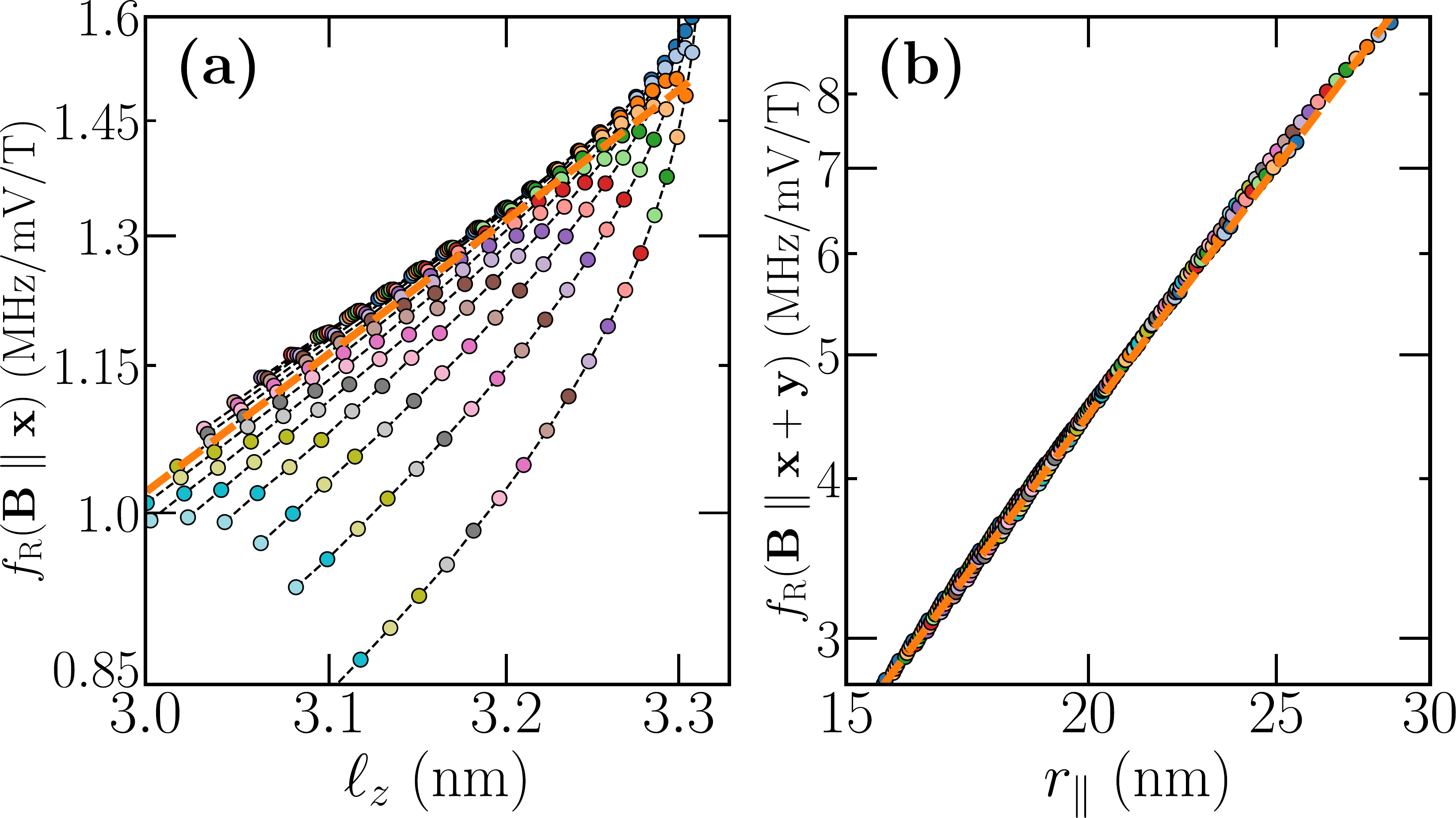}
    \caption{(a) Correlations between the Rabi frequency $f_\mathrm{R}(\mathbf{B}\parallel\mathbf{x})$ at constant magnetic field and the dot size $\ell_z$, collected from the map of Fig.~\ref{fig:Vgmap}a (opposite drives on the L and R gates, NS mechanism). Dotted black lines connect points at the same $V_\mathrm{C}$, while symbol colors label points at the same $V_\mathrm{bg}$. The dashed orange line is a guide to the eye with slope $s=4$. (b) Same for $f_\mathrm{R}(\mathbf{B}\parallel\mathbf{x}+\mathbf{y})$ {\it vs} dot size $r_\parallel$ and a drive on the L gate only, collected from Fig.~\ref{fig:VgmapTMR} (conventional $g$-TMR plus NS mechanism). The dashed orange line is a guide to the eye with slope $s=2$.}
    \label{fig:correlations}
\end{figure}

The non-separability of the electrical \YMN{confinement potential $V(\mathbf{r})$} in the present device is highlighted in Fig.~\ref{fig:potlandscape}a-c. The vertical electric field $E_z=-\partial V/\partial z$ does, in particular, show a significant dependence on $x$. The non-separability is promoted by the thick top barrier where the electric field lines connecting the gates engulf. \YMN{The AC potential $V_{\rm ac}(\mathbf{r})$ is also non-separable, which makes an additional contribution to $\gt_{zx}^\prime$ (see Appendix \ref{app:gTMR}); however the Rabi frequencies calculated for the homogeneous, average AC electric field $E_{\mathrm{ac},x}\approx 1.7$\,$\mu$V/nm are comparable to Fig.~\ref{fig:Angularmap}, which suggests that the Rabi oscillations are  dominated by the non-separability of the confinement potential in the present device.} 

The Rabi frequencies $f_\mathrm{R}(\mathbf{B}\parallel\mathbf{x})$ (NS mechanism) and $f_\mathrm{R}(\mathbf{B}\parallel\mathbf{z})$ (cubic Rashba SOI) are plotted as a function of the front gate voltage $V_\mathrm{C}$ and back gate voltage $V_\mathrm{bg}$ in Fig.~\ref{fig:Vgmap}. The NS mechanism prevails over cubic Rashba SOI at almost any bias. Increasingly negative $V_\mathrm{C}$ enhances both radial and vertical electric fields, which reduces the radius $r_\parallel$ as well as the vertical extension $\ell_z=\sqrt{\langle z^2\rangle-\langle z\rangle^2}$ of the dot. Increasingly positive $V_\mathrm{bg}$ mostly enhances the vertical electric field and decreases $\ell_z$ (and, to a much lesser extent than the front gate, $r_\parallel$). The maximum vertical electric field is, however, limited by the formation of a triangular well at the top SiGe/Al$_2$O$_3$ interface, which captures the hole \cite{Su17} in the gate voltage space outlined by the white areas. The strength of the cubic Rashba SOI is proportional to small vertical electric fields, but saturates then decreases at large ones \cite{Wang21}, and the resulting Rabi frequency is expected to scale as $r_\parallel^2$ \cite{Terrazos21}. Therefore, $f_\mathrm{R}(\mathbf{B}\parallel\mathbf{z})$ is optimal at moderate $V_\mathrm{bg}$, and decreases with increasing $V_\mathrm{C}<0$. On the contrary, the NS mechanism responsible for $f_\mathrm{R}(\mathbf{B}\parallel\mathbf{x})$ subsides with increasing $V_\mathrm{bg}>0$ (decreasing $\ell_z$) but shows little correlations with $r_\parallel$. In the standard situation where the confinement is much stronger along $z$ than in the $xy$ plane, the non-separable motion of the hole is, indeed, mostly limited by the electrical polarizability along $z$, hence by the $\propto 1/\ell_z^2$ gap between HH states. Moreover, the HH/LH gap $\Delta_\mathrm{LH}\approx(2\pi^2\hbar^2\gamma_2)/(m_0L_\mathrm{W}^2)+2(\nu+1)b_v\varepsilon_\parallel$ is the sum of a $\propto 1/L_\mathrm{W}^2$ structural confinement energy in the well and a $\propto\varepsilon_\parallel$ strain-induced splitting (with $b_v$ the uniaxial deformation potential of the valence band, $\nu$ the Poisson ratio of Ge and $\varepsilon_\parallel$ the in-plane strain in the well). This gap thus also tends to increase with vertical field (due to the decrease of the effective width of the well $L_\mathrm{W}^\mathrm{eff}\propto\ell_z$), which is however concealed here by the large contribution of strains ($\varepsilon_{\parallel}=-0.63\%$). Finally, the non-separability becomes less relevant when the dot thins down, so that $f_\mathrm{R}(\mathbf{B}\parallel\mathbf{x})$ scales altogether as $\ell_z^4$ at constant magnetic field (see Appendix \ref{app:theory}). This $\ell_z^4$ scaling is prominent in Fig.~\ref{fig:correlations}a when increasing $V_\mathrm{bg}>0$. When increasing $V_\mathrm{C}<0$, it is superseded by the front-gate voltage dependence of the non-separable electric field patterns within the dot. At constant Larmor frequency, $f_\mathrm{R}(\mathbf{B}\parallel\mathbf{x})$ may also increase with increasingly negative $V_\mathrm{C}$ due to the concomitant decrease of $\gt_x$, as shown in Fig.~\ref{fig:drives}a.

\section{Rabi oscillations with an asymmetric drive}

\begin{figure}
    \centering
    \includegraphics[width=1\columnwidth]{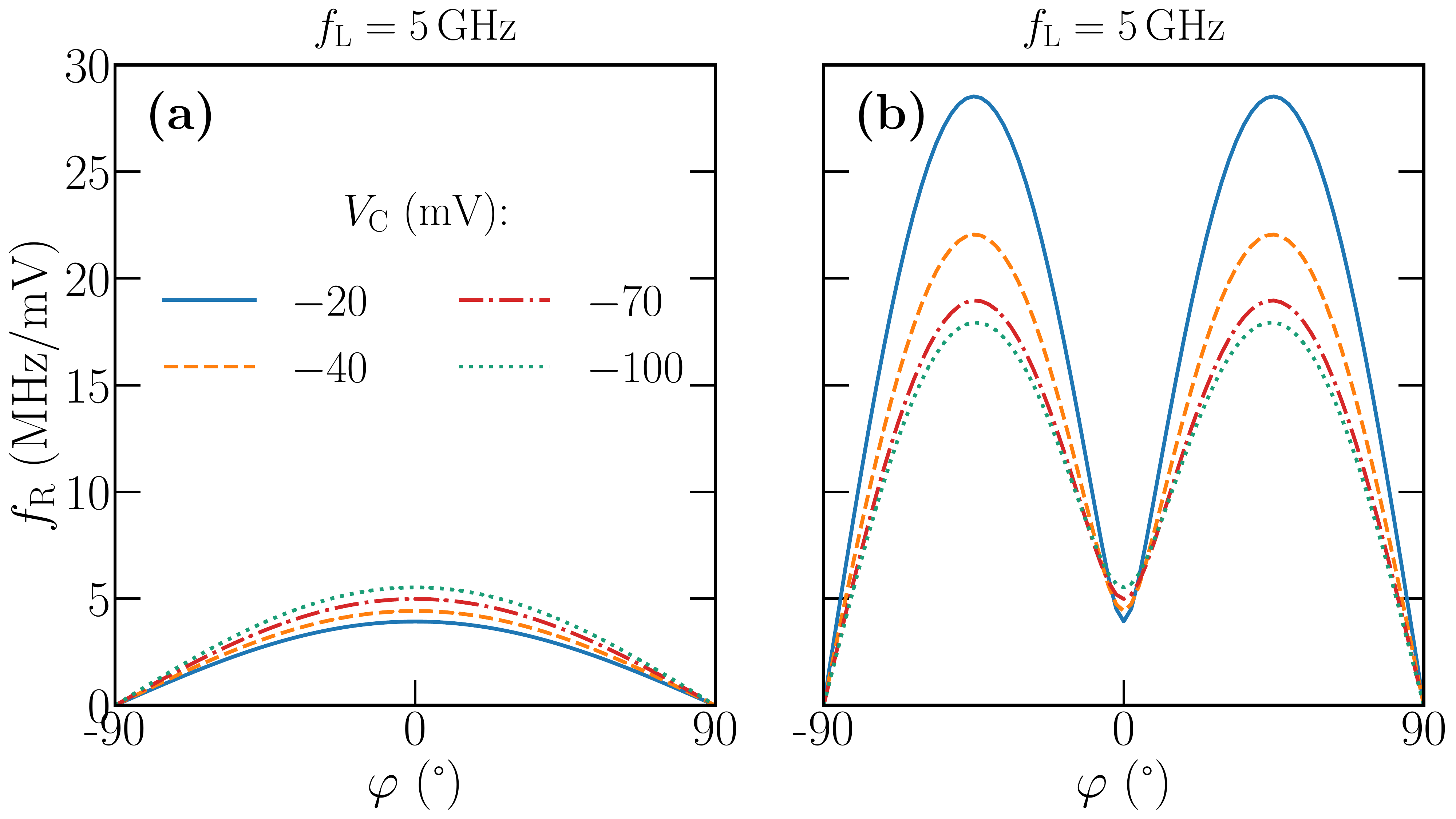}
    \caption{Rabi frequency for an in-plane magnetic field ($\theta=90^\circ)$ as a function of $\varphi$, for constant $f_\mathrm{L}=5$\,GHz and different $V_\mathrm{C}$'s ($V_\mathrm{bg}=0$\,V). The hole is driven with opposite modulations $\delta V_\mathrm{L}=-\delta V_\mathrm{R}=(V_\mathrm{ac}/2)\cos\omega_\mathrm{L}t$ on the L and R gates. (b) Same for a hole driven by a modulation $\delta V_\mathrm{L}=V_\mathrm{ac}\cos\omega_\mathrm{L}t$ on the L gate only.}
    \label{fig:drives}
\end{figure}

\begin{figure}
    \centering
    \includegraphics[width=1\columnwidth]{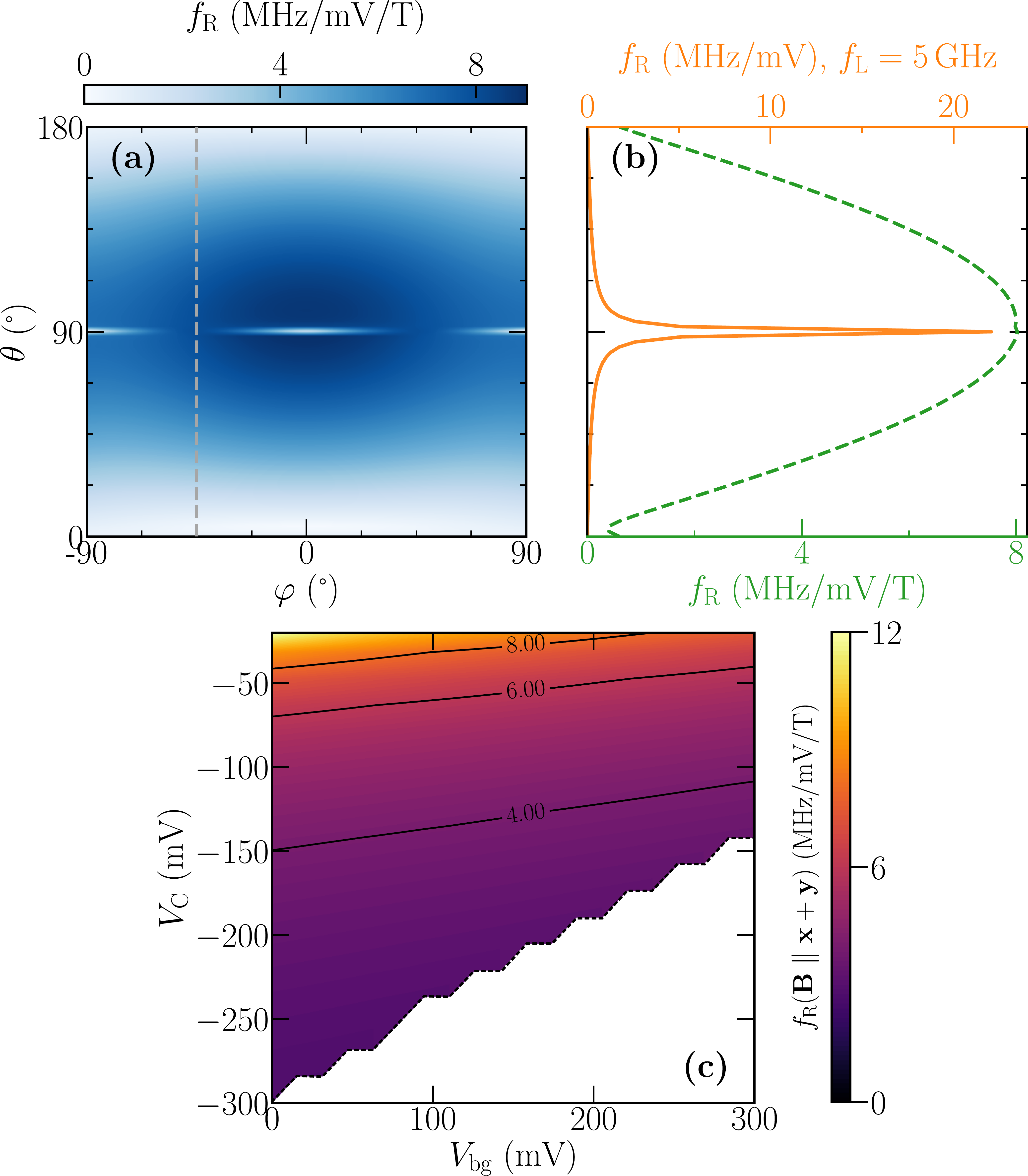}
    \caption{(a) Map of the Rabi frequency $f_\mathrm{R}$ as a function of the magnetic field angles $\theta$ and $\varphi$ defined in Fig.~\ref{fig:device}, for a drive $\delta V_\mathrm{L}=V_\mathrm{ac}\cos\omega_\mathrm{L}t$ on the L gate only ($V_\mathrm{C}=-40$\,mV, $V_\mathrm{bg}=0$\,V, $V_\mathrm{ac}=1$\,mV and $B=1$\,T). (b) Cut along the dashed gray line in (a), at constant magnetic field $B=1$\,T (green), and at constant Larmor frequency $f_\mathrm{L}=\omega_\mathrm{L}/(2\pi)=5$\,GHz (orange). (c) Map of the Rabi frequency $f_\mathrm{R}(\mathbf{B}\parallel\mathbf{x}+\mathbf{y})$ at constant magnetic field $B=1$\,T as a function of $V_\mathrm{C}$ and $V_\mathrm{bg}$ (drive on the L gate only). The hole is pulled by the electric field from the well to the top SiGe/Al$_2$O$_3$ interface in the white areas.}
    \label{fig:VgmapTMR}
\end{figure}

We may, alternatively, drive the hole with a single gate (as is practically done in most experiments \cite{Scappucci20,Hendrickx20b,Hendrickx20,Hendrickx21}). We emphasize, though, that direct manipulation with the C gate is particularly inefficient with an in-plane magnetic field for symmetry reasons (see Appendix \ref{app:central}) \cite{Venitucci18}. The Rabi frequency resulting from a modulation $\delta V_\mathrm{L}=V_\mathrm{ac}\cos\omega_\mathrm{L}t$ on the L gate only is plotted as a function of the angle $\varphi$ of an in-plane magnetic field in Fig.~\ref{fig:drives}b (at constant $f_\mathrm{L}=5$\,GHz). The NS mechanism is still responsible for the Rabi oscillations when $\mathbf{B}\parallel\mathbf{x}$ ($\varphi=0$). It is, however, superseded by two prominent peaks at $\varphi=\pm 45^\circ$. The latter arise from direct modulations of the principal $\gt$-factors $\gt_x$ and $\gt_y$ (``conventional'' $g$-TMR) by the inhomogeneous AC field of the L gate that squeezes the dot dynamically. This is highlighted in Fig.~\ref{fig:potlandscape}e-f, which shows how the AC field acts on the left side while being inefficient on the right side of the dot. When the dot is driven with opposite modulations on the L and R gates as in the previous section, the AC field is actually more homogeneous, but above all too symmetric to modulate the principal $\gt$-factors \cite{Venitucci18}. \YMN{As shown in Appendix \ref{app:gTMR}, conventional $g$-TMR results, at lowest order in $1/\Delta_\mathrm{LH}$, from the interplay between the drive, $R$ and $H_\mathrm{Z}$, or $R$ and $S$}, in contrast with the NS mechanism (which involves $S$ and $H_\mathrm{Z}$). According to Eq.~(\ref{eq:fRg}), the resulting contribution to the Rabi frequency is $\propto(\gt_x^\prime\YMN{+}\gt_y^\prime)b_xb_y$ for an in-plane magnetic field and therefore vanishes when $\mathbf{B}\parallel\mathbf{x}$ or $\mathbf{B}\parallel\mathbf{y}$. The complete dependence of the Rabi frequency on the magnetic field orientation, and the map of $f_\mathrm{R}(\mathbf{B}\parallel\mathbf{x}+\mathbf{y})$ as a function of $V_\mathrm{C}$ and $V_\mathrm{bg}$ are plotted in Fig.~\ref{fig:VgmapTMR}. At constant magnetic field, conventional $g$-TMR is in fact optimal when $\mathbf{B}$ makes a finite angle with the $xy$ plane \cite{Venitucci19,Michal21}; at constant Larmor frequency, it is however optimal for $\mathbf{B}\parallel\mathbf{x}\pm\mathbf{y}$ owing to the large $\gt_z/\YMN{|}\gt_{x,y}\YMN{|}$ ratio. The Rabi frequency shows a strong dependence on the front gate voltage $V_\mathrm{C}$, which results from an intrinsic $\propto r_\parallel^2$ scaling (see Fig.~\ref{fig:correlations}b and Appendix \ref{app:gTMR}). There is little dependence on the back gate voltage $V_\mathrm{bg}$ (hence on $\ell_z$). 

\section{Discussion}

\YMN{We would first like to emphasize that the NS mechanism essentially results in the present device from the non-separability of the confinement potential while conventional $g$-TMR results from the inhomogeneity and asymmetry of the drive. We also underline that the confinement potential would be separable if harmonic [$V_\mathrm{conf}(x,y,z)\approx m_\parallel\omega_\parallel^2(x^2+y^2)/2+m_\perp\omega_\perp^2(z-z_0)^2/2$ given the symmetries of the device]. Therefore, non-separability implies significant anharmonicity within the dot (but anharmonicity does not conversely imply non-separability of the confinement potential). Anharmonicity had already been identified as an other possible enabler of conventional $g$-TMR \cite{Venitucci18,Venitucci19,Salfi22}; anharmonicity and/or drive inhomogeneity therefore appear as general ingredients of $g$-TMR-like mechanisms.}

\begin{figure}
    \centering
    \includegraphics[width=1\columnwidth]{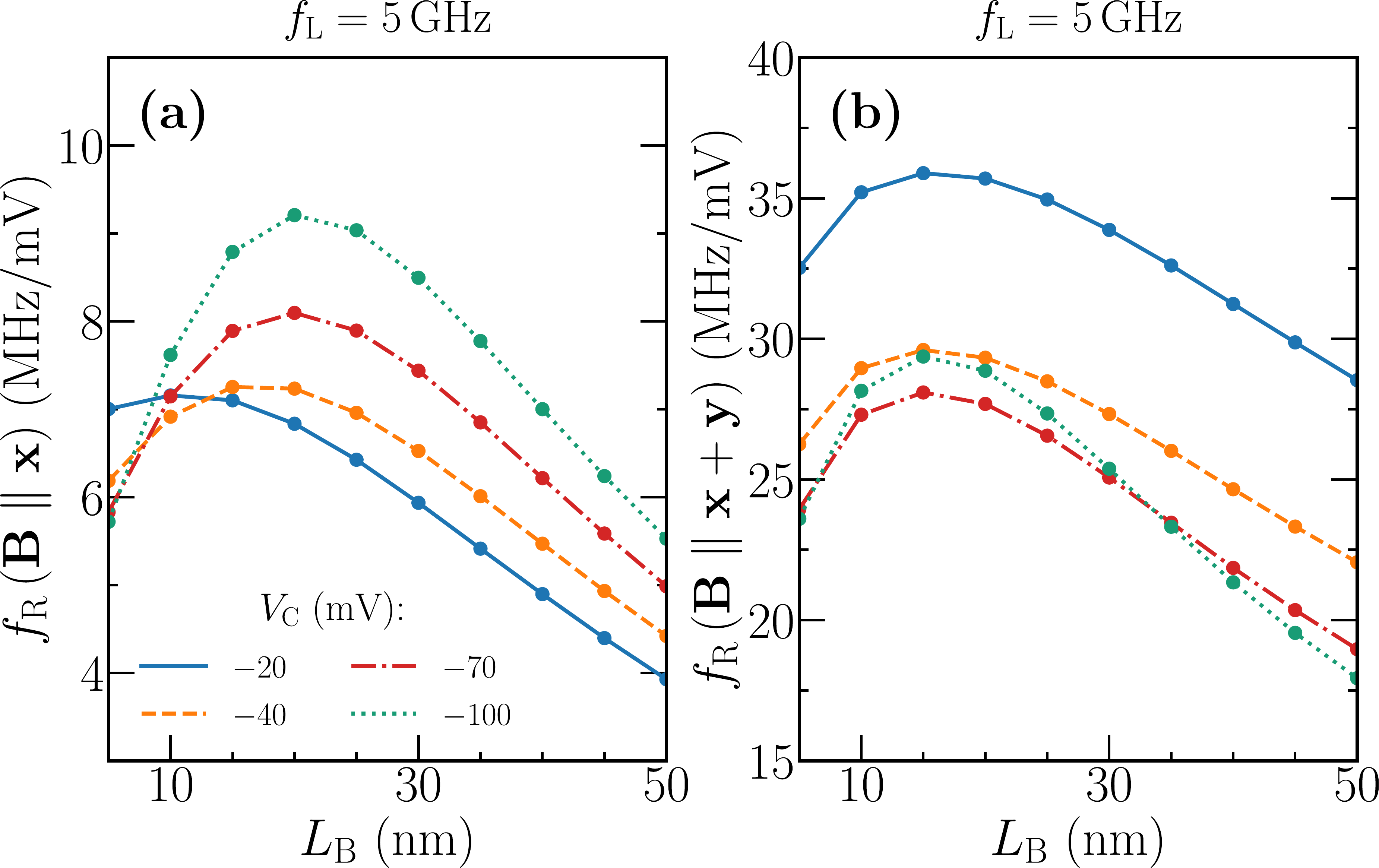}
    \caption{(a) Rabi frequency $f_\mathrm{R}(\mathbf{B}\parallel\mathbf{x})$ at constant $f_\mathrm{L}=5$\,GHz as a function of the upper barrier thickness $L_\mathrm{B}$ for different $V_\mathrm{C}$'s and $V_\mathrm{bg}=0$\,V (opposite drives on L and R gates, NS mechanism). (b) Same for $f_\mathrm{R}(\mathbf{B}\parallel\mathbf{x}+\mathbf{y})$ (drive on the L gate only, conventional $g$-TMR plus NS mechanism).}
    \label{fig:drives20}
\end{figure}

The Rabi frequencies tend to increase with decreasing upper barrier thickness $L_\mathrm{B}$ as a result, in part, of the enhancement of the AC electric field from the closer lateral gates. At small enough $L_\mathrm{B}$ however, the electric field is screened by the C gate and the Rabi frequencies decrease again. Therefore, the Rabi frequencies are typically optimal around $L_\mathrm{B}\simeq 20$ nm (see Fig.~\ref{fig:drives20}).

The above conventional $g$-TMR and NS mechanisms are quantitatively, but not qualitatively affected if the dot is shifted from the center of the C gate by a static positive bias on one of the side gates. The resulting mild symmetry breaking may lift the exact zeros of Fig.~\ref{fig:drives} at $\varphi=\pm 90^\circ$; $\YMN{|}\gt_x\YMN{|}$ and $\YMN{|}\gt_y\YMN{|}$ also become different, which mostly impacts the anisotropy of the Rabi frequency at constant $f_\mathrm{L}$. \YMN{Moreover, $\gt_x^\prime$ and $\gt_y^\prime$ can now be non-zero for opposite drives on the L and R gates, so that the map of Fig.~\ref{fig:Angularmap}a acquires a similar $g$-TMR background as Fig.~\ref{fig:VgmapTMR}a.} Another interesting operating mode would be to squeeze the dot laterally with positive voltages on, e.g., both B and T gates, and drive with opposite modulations on the L and R gates, in order to leverage the strong linear Rashba SOI emerging in the limit $\ell_x\gg\ell_y\sim\ell_z$ \cite{Bosco21b}. It is, however, practically difficult to achieve $\ell_y/\ell_z\lesssim 4$ with the setup of Fig.~\ref{fig:device}, because the electrical in-plane confinement is much softer than the structural vertical confinement (even with oval-shaped gates), and because the hole tends to be pulled out from the well to the surface as the action of the repulsive B and T gates is screened there by the C gate. Moreover, the emerging linear Rashba SOI interferes destructively with the NS mechanism when the magnetic field lies in-plane, which further hinders its exploitation. 

A linear Rashba SOI may also emerge from symmetry breaking at the Ge/GeSi interface \cite{Xiong21,Liu22}. This interaction is not described by the present LK Hamiltonian and can only be captured by atomistic methods. It increases with increasing vertical electric field, and the resulting Rabi frequency scales as $r_\parallel^4$. According to the estimates of Ref.~\cite{Liu22} (however drawn in Ge/Si rather than Ge/GeSi superlattices), this Rabi frequency shall be of the order of \YMN{4.3\,MHz/mV} at $f_\mathrm{L}=5$\,GHz ($\mathbf{B}\parallel\mathbf{x}$) in a vertical electric field $E_z=30$\,kV/cm (above which the hole is pulled out from the well), for a dot radius $r_\parallel=27$ nm (consistent with the wave function of Fig.~\ref{fig:device}) \footnote{The dot is driven by opposite AC modulations on the L and R gates.}. It is, therefore, of the same order of magnitude as the $g$-TMR Rabi frequencies discussed in this work. The actual strength of this interaction for a disordered GeSi barrier (which may moreover sligthly interdiffuse with the Ge well) remains, however, to be assessed accurately. Additional research is hence needed to make the most of the different flavours of SOI appearing in semiconductor heterostructure hole spin qubit devices.

\section{Conclusions}

To conclude, we have discussed the manipulation of heavy-hole spin qubits in planar semiconductor heterostructures (considering Ge/SiGe as an illustration). We have shown that for realistic, yet highly patterned electric field distributions, the Rabi oscillations can be dominated by $g$-TMR-like mechanisms resulting from modulations of the shape of the dot rather than by Rashba SOI. In particular, the coupling between the in-plane and out-of-plane motions of the hole in the \YMN{non-separable potential of the gates can give rise to sizable Rabi frequencies}. This non-separability (NS) mechanism, which has been overlooked in previous studies, is prevalent for in-plane magnetic fields, and for small vertical electric fields where the hole can best move out-of-plane. Since the in-plane $\gt$-factors of heavy holes are small, it can outweigh cubic Rashba SOI by orders of magnitude at a given Larmor frequency. The NS mechanism can be superseded by direct modulations of the principal $\gt$-factors when the driving electric field is \YMN{sufficiently asymmetric}, as is usually the case when the hole is manipulated with a single nearby gate. This collection of $g$-TMR mechanisms shall therefore play a role in the experiments of Refs.~\cite{Hendrickx20,Hendrickx21}. Their fingerprints could actually be revealed in a complete experimental map of the Rabi frequencies as a function of the orientation of the magnetic field \cite{Crippa18}. In general, $g$-TMR provides opportunities for spin manipulation with in-plane magnetic fields where heavy holes best decouple from hyperfine noise. The NS mechanism is, in particular, purely transverse for such in-plane magnetic fields, and thus does not enhance the sensitivity of the hole to electrical dephasing noise. A reliable exploitation may, however, call for a careful engineering of the $\gt$-factors of the dots (through, e.g., strain optimization \cite{Lodari22}), as the constraints on magnetic field alignment are the more stringent as the $\gt$-factors anisotropy is large. This work also highlights that a comprehensive understanding of hole spin qubits in semiconductor heterostructures calls for a detailed modeling of these systems going beyond simple models and assumptions. The design and optimization of spin qubits devices shall therefore benefit from microscopic modeling able to cope with their complexity.

\section*{Acknowledgements}

We thank M. Filippone and V. Michal for fruitful discussions and comments on the manuscript. This work was supported by the French National Research Agency (ANR) through the MAQSi project and the PEPR project PRESQUILE.


\appendix

\section{Hamiltonian and $\gt$-factors of a heavy-hole doublet}
\label{app:hamiltonian}

In this appendix, we discuss the Hamiltonian of the holes as well as the $\gt$-factors of the ground-state heavy-hole doublet.

We consider a hole moving in a potential $V(\mathbf{r})$ and in a homogeneous magnetic field $\mathbf{B}$. We apply hard-wall boundary conditions outside a quantum well with thickness $L_\mathrm{W}$ [$V(\mathbf{r})=+\infty$ if $|z|\ge L_\mathrm{W}/2$]. The motion of the hole can be solved with the four bands Luttinger-Kohn (LK) Hamiltonian \cite{Luttinger56,KP09}. The latter describes the coupled dynamics of the heavy and light-hole components of the wave function, mapped respectively onto the the $J_z=\pm\tfrac{3}{2}$ and $J_z=\pm\tfrac{1}{2}$ components of a $J=\tfrac{3}{2}$ spin. In the $\{\ket{\tfrac{3}{2},\tfrac{3}{2}},\ket{\tfrac{3}{2},\tfrac{1}{2}},\ket{\tfrac{3}{2},-\tfrac{1}{2}},\ket{\tfrac{3}{2},-\tfrac{3}{2}}\}$ basis set, the kinetic LK Hamiltonian reads:
\begin{equation}
    H_\mathrm{LK}=\begin{pmatrix}
    P+Q & -S & R & 0 \\
    -S^\dagger & P-Q & 0 & R \\
    R^\dagger & 0 & P-Q & S \\
    0 & R^\dagger & S^\dagger & P+Q
    \end{pmatrix}
    \label{eq:LK}
\end{equation}
where: 
\begin{subequations}
\begin{align}
P&=\frac{1}{2m_0}\gamma_1(p_x^2+p_y^2+p_z^2) \\
Q&=\frac{1}{2m_0}\gamma_2(p_x^2+p_y^2-2p_z^2) \\
R&=\frac{1}{2m_0}\sqrt{3}\left[-\gamma_2(p_x^2-p_y^2)+2i\gamma_3p_xp_y\right] \\
S&=\frac{1}{2m_0}2\sqrt{3}\gamma_3(p_x-ip_y)p_z\
\end{align}
\end{subequations}
with $\mathbf{p}$ the momentum and $\gamma$'s the Luttinger parameters that characterize the mass of the hole. Note that we assume here holes with positive (electron-like) dispersion. At finite magnetic field $\mathbf{B}$, the heavy and light-hole components are also split and mixed by the Zeeman Hamiltonian:
\begin{equation}
H_\mathrm{Z}=2\mu_B(\kappa\mathbf{B}\cdot\mathbf{J}+q\mathbf{B}\cdot\mathbf{J}^3)\,,\label{eq:zeeman}
\end{equation}
where $\mathbf{J}$ is the spin $\tfrac{3}{2}$ operator, $\mathbf{J}^3\equiv(J_x^3,J_y^3,J_z^3)$, and $\kappa$, $q$ are the isotropic and cubic Zeeman parameters. The $\mathbf{J}$ matrices consistent with the basis set of Eq.~(\ref{eq:LK}) read:
\begin{subequations}
\begin{align}
J_x&=\frac{1}{2}\begin{pmatrix}
0 & \sqrt{3} & 0 & 0  \\
\sqrt{3} & 0 & 2 & 0  \\
0 & 2 & 0 & \sqrt{3}  \\
0 & 0 & \sqrt{3} & 0 
\end{pmatrix} \\
J_y&=\frac{i}{2}\begin{pmatrix}
0 & -\sqrt{3} & 0 & 0  \\
\sqrt{3} & 0 & -2 & 0 \\
0 & 2 & 0 & -\sqrt{3} \\
0 & 0 & \sqrt{3} & 0 \\
\end{pmatrix} \\
J_z&=\frac{1}{2}\begin{pmatrix}
3 & 0 & 0 & 0 \\
0 & 1 & 0 & 0 \\
0 & 0 & -1 & 0 \\
0 & 0 & 0 & -3 \\
\end{pmatrix}\,.
\end{align}
\end{subequations}
The action of the magnetic field on the orbital motion of the hole is described by the substitution $\mathbf{p}\to-i\hbar\boldsymbol{\nabla}+e\mathbf{A}$, where $\mathbf{A}=\mathbf{B}\times\mathbf{r}/2$ is the vector potential \footnote{\YMN{The LK Hamiltonian must first be symmetrized: $p_ip_j\to(p_ip_j+p_jp_i)/2$.}}. Finally, biaxial strains in the well may further split the heavy and light holes by an energy $\Delta_\mathrm{BP}=2(\nu+1)b_v\varepsilon_\parallel$, where $\varepsilon_\parallel$ is the in-plane strain, $\nu$ the Poisson ratio and $b_v$ the uniaxial valence band deformation potential of the host material. The total Hamiltonian is therefore
\begin{equation}
H_0=H_\mathrm{LK}+H_\mathrm{Z}+V(\mathbf{r})-(\nu+1)b_v\varepsilon_\parallel J_z^2
\end{equation}
with hard-wall boundary conditions at $z=\pm L_\mathrm{W}/2$.

In the absence of HH/LH mixing ($R=S=0$), the eigenstates of $H_0$ at $\mathbf{B}=\mathbf{0}$ are pure heavy-hole ($J_z=\pm\tfrac{3}{2}$) and pure light-hole ($J_z=\pm\tfrac{1}{2}$) doublets. They are separated by (at least) the HH/LH band gap 
\begin{equation}
\Delta_\mathrm{LH}=\frac{2\pi^2\hbar^2\gamma_2}{m_0L_\mathrm{W}^2}+2(\nu+1)b_v\varepsilon_\parallel\,,
\end{equation}
where the first term is the splitting between the ground-state HH and LH subbands due to the vertical confinement in the quantum well, and the second term is the splitting due to the biaxial strains \cite{Michal21}. \YMN{These doublets are further split at finite magnetic field by the Zeeman Hamiltonian $H_\mathrm{Z}$; each can be characterized by an effective Hamiltonian ${\cal H}=\tfrac{1}{2}\mu_B\boldsymbol{\sigma}\cdot\gt\mathbf{B}$, where $\gt$ is the gyromagnetic $\gt$-matrix of the doublet. For heavy holes, $\gt$ is diagonal in the $\{\ket{\tfrac{3}{2}},\,\ket{-\tfrac{3}{2}}\}$ basis set with elements $\gt_x=3q$, $\gt_y=-3q$ and $\gt_z=6\kappa+27q/2$.}

\YMN{The $R$ and $S$ terms of the LK Hamiltonian actually admix heavy and light holes. The renormalized $\gt$-matrices can be obtained by a Schrieffer-Wolff transformation that integrates out the LH degrees of freedom \cite{Michal21}; assuming for now a fully separable potential $V(x,y,z)$, the principal $\gt$-factors of the ground-state HH doublet read, to order $1/\Delta_\mathrm{LH}$:}
\begin{subequations}
\label{eq:gHH}
\begin{align}
\gt_x&=+3q+\frac{6}{m_0\Delta_\mathrm{LH}}\big[\kappa\gamma_2\left(\langle p_x^2\rangle-\langle p_y^2\rangle\right) \nonumber \\
&-2\eta_h\gamma_3\left(\gamma_3\langle p_x^2\rangle-\gamma_2\langle p_y^2\rangle\right)\big] \\
\gt_y&=\YMN{-}3q\YMN{-}\frac{6}{m_0\Delta_\mathrm{LH}}\big[\kappa\gamma_2\left(\langle p_y^2\rangle-\langle p_x^2\rangle\right) \nonumber \\
&-2\eta_h\gamma_3\left(\gamma_3\langle p_y^2\rangle-\gamma_2\langle p_x^2\rangle\right)\big] \\
\gt_z&=6\kappa+\frac{27}{2}q-2\gamma_h+\delta\mathsf{g}_z\,,
\end{align}
\end{subequations}
\YMN{where $\gamma_h$, $\eta_h$ are dimensionless parameters that depend on vertical confinement, defined in Ref.~\cite{Michal21}. In unstrained Ge films with hard wall boundary conditions, $\gamma_h\approx 3.56$ and $\eta_h\approx 0.20$ whatever $L_\mathrm{W}$; in the present biaxial strains $\gamma_h\approx 2.59$ and $\eta_h\approx 0.42$ for $L_\mathrm{W}=16$\,nm}. The expectations values of $p_x$ and $p_y$ are calculated for the ground-state heavy-hole envelope of the quantum dot at $\mathbf{B}=\mathbf{0}$, and $\delta\gt_z$ collects corrections of order $\langle p_x^2\rangle/(m_0\Delta_\mathrm{LH})$ and $\langle p_y^2\rangle/(m_0\Delta_\mathrm{LH})$ \cite{Katsaros11}. The $\propto\kappa\gamma_2$ contributions to $\gt_x$ and $\gt_y$ result from the interplay between the Zeeman Hamiltonian $H_\mathrm{Z}$ and the $R$ term of the LK Hamiltonian, while the $\propto\eta_h$ terms result from the action of the magnetic field on the orbital motion of the holes through the substitution $\mathbf{p}\to-i\hbar\boldsymbol{\nabla}+e\mathbf{A}$ in $R$ and the interplay with $S$.

In a circular quantum dot, $\langle p_x^2\rangle=\langle p_y^2\rangle$, so that $\gt_x=\YMN{-}\gt_y\ll\gt_z$. The deviations from $\gt_x=\YMN{-}\gt_y=3q$ moreover scale as $\langle p_{x,y}^2\rangle\propto\eta_h/r_\parallel^2$ \cite{Wang22}, where $r_\parallel=\sqrt{\langle x^2+y^2\rangle}$ is the radius of the dot. \YMN{They arise from the orbital motion of the hole in the magnetic vector potential.}

\section{Heavy-hole Hamiltonian in a non-separable confinement potential}
\label{app:theory}

In this appendix, we discuss the effective low-energy Hamiltonian of a heavy hole in a heterostructure with non-separable confinements in the $xy$ plane and along the growth direction $z$. We show that the non-separability of the confinement potential gives rise to a specific $g$-TMR mechanism. \YMN{The effects of the non-separability of the AC electric field will be discussed in Appendix \ref{app:gTMR}.}

We consider as in Appendix \ref{app:hamiltonian} a quantum well with thickness $L_\mathrm{W}$ subject to the model confinement potential:
\begin{equation}
\label{eq:potential}
V_\mathrm{conf}(x,y,z)=\frac{1}{2}m_0\omega_0^2(x^2+y^2)\zeta(z)^2.
\end{equation}
This potential is harmonic with respect to $x$ and $y$, with a characteristic in-plane confinement energy $\hbar\omega_0$ modulated by the function $\zeta(z)$. It accounts for the main features of the $g$-TMR mechanism discussed here. In particular, we do not need to introduce an extra, static vertical electric field in the model to leverage the SOI. We neglect in the following the action of the magnetic vector potential on the orbital motion of the holes (it is accounted for in the numerical simulations but is not central to our argument).

We start the analysis from the uncoupled heavy and light-hole states ($R=S=0$) and deal with the HH/LH mixings in perturbation. The heavy and light-hole energies and wave functions cannot, however, be solved exactly for an arbitrary function $\zeta(z)$. In the following, we therefore make a reasonable ansatz for these wave functions, based on a minimal modification of the solutions of the separable harmonic potential problem. For that purpose, it is convenient to introduce the harmonic lengths $\ell_{h,l}=\sqrt{\hbar/(m^{h,l}_\parallel\omega_{h,l})}$ of the heavy and light holes, where $\omega_{h,l}/\omega_0=\sqrt{m_0/m^{h,l}_\parallel}$, and $m_\parallel^{h,l}=m_0/(\gamma_1\pm\gamma_2)$ are the in-plane hole masses. The effect of the non-separability function $\zeta(z)$ is then accounted for by the substitution $\ell_{h,l}\rightarrow \ell_{h,l}(z)=\ell_{h,l}/\sqrt{\zeta(z)}$ in the in-plane harmonic oscillator wave functions:
\begin{align}
    \psi_{n}^{h,l}(r_i,z)&=\frac{1}{2^nn!}\sqrt[4]{\frac{1}{\pi \ell_{h,l}^2(z)}} \nonumber \\
    &\times\exp\left(-\frac{r_i^2}{2\ell_{h,l}^2(z)}\right)H_{n}\left(\frac{r_i}{\ell_{h,l}(z)}\right)\,, 
    \label{eq:wavefun}
\end{align}
where $r_i\in\{x,y\}$ is an in-plane coordinate, $n\ge 0$ is a quantum number and $H_{n}$ is the associated Hermite polynomial. We hence write the total heavy and light-hole wave functions as:
\begin{equation}
    \Psi_{n_x,n_y,n_z}^{h,l}(x,y,z)=\psi_{n_x}^{h,l}(x,z)\psi_{n_y}^{h,l}(y,z)Z_{n_z}(z)\,,
    \label{eq:wavefun2}
\end{equation}
where:
\begin{equation}
    Z_{n_z}(z)=\sqrt{\frac{2}{L_\mathrm{W}}}\sin\left(\frac{(n_z+1)\pi(L_\mathrm{W}/2+z)}{L_\mathrm{W}}\right)\,.
\end{equation} 
Despite the dependence of the harmonic lengths on $z$, the above wave functions remain normalized and fulfill the expected orthogonality relations. This ansatz should be a good approximation to the exact wave functions as long as $\zeta(z)$ does not deviate much from 1 in the well \footnote{Alternatively, it is possible to reach the same conclusions without making this ansatz but going to fourth order in perturbation theory.}.
For simplicity, we introduce the ket notation for the spinor wave functions:
\begin{subequations}
\begin{align}
    \ket{n_x,n_y,n_z,J_z=\pm\tfrac{3}{2}}&=\ket{\Psi_{n_x,n_y,n_z}^h}\otimes\ket{\tfrac{3}{2},J_z=\pm\tfrac{3}{2}} \\
    \ket{n_x,n_y,n_z,J_z=\pm\tfrac{1}{2}}&=\ket{\Psi_{n_x,n_y,n_z}^l}\otimes\ket{\tfrac{3}{2},J_z=\pm\tfrac{1}{2}}\,.
\end{align}
\label{eq:hhlhbasis}
\end{subequations}
 
We next add an AC electric field $\mathbf{F}_\mathrm{ac}(t)$ in order to drive the hole spin. We set $\mathbf{F}_\mathrm{ac}(t)=\mathbf{F}\cos(\omega_\mathrm{d}t)$ where $\mathbf{F}=(F_x,F_y,0)$ lies in-plane. The driving Hamiltonian is therefore $H_\mathrm{d}(t)=-e\mathbf{F}\cdot\mathbf{r}\cos(\omega_\mathrm{d}t)$. We find that such a separable drive is sufficient to achieve $\gt$-TMR in a non-separable confinement potential $V_\mathrm{conf}$. Intuitively, the displacement of the hole under the drive depends on how tight is the lateral confinement. The harmonic confinement strength, $\propto m_0\omega_0^2$ in the separable case, is here modulated along $z$ by the function $\zeta(z)^2$. In striking contrast with the separable case, the dot is therefore inhomogeneously displaced (sheared) by the drive.

Since we are only interested on the effect of the drive onto the heavy-hole ground-state subspace $\ket{0,0,0,\pm\tfrac{3}{2}}$, we can make use of a Schrieffer-Wolff transformation to integrate out the couplings with the excited states. We find that the drive induces a spin-dependent response at third order in the perturbation series. The third-order correction to the heavy-hole Hamiltonian is \cite{Winkler03}
\begin{align}
    H^{(3)}_{m,m'}=-\frac{1}{2}\sum_{l,m''}\bigg[&\frac{H'_{ml}H'_{lm''}H'_{m''m'}}{(E_{m'}-E_l)(E_{m''}-E_l)}+ \nonumber \\ &\frac{H'_{mm''}H'_{m''l}H'_{lm'}}{(E_{m}-E_l)(E_{m''}-E_l)}\bigg] \nonumber \\
    +\frac{1}{2}\sum_{l,l'}H'_{ml}H'_{ll'}H'_{l'm'}\bigg[&\frac{1}{(E_m-E_l)(E_m-E_l')}+ \nonumber \\
    &\frac{1}{(E_m'-E_l)(E_m'-E_l')}\bigg]\,,
    \label{eq:SW3}
\end{align}
\YMN{where $m$, $m'$, and $m''$ run over the HH ground-state subspace, $l$ and $l'$ run over the \YMN{HH and LH} excitations [Eqs.~(\ref{eq:hhlhbasis})]}, $E_i$ are the bare energies and $H'$ is the Hamiltonian coupling both subspaces. \YMN{In the present case, $H'$ collects the off-diagonal terms of the LK and Zeeman Hamiltonians for LH excitations [Eqs.~(\ref{eq:LK}) and (\ref{eq:zeeman})], and the $\propto\mathbf{F}\cdot\mathbf{r}$ drive field for HH excitations. Hence, the relevant drive contributions to the effective ground-state Hamiltonian, are, to order $1/\Delta_\mathrm{LH}$, proportional to:}
\begin{align}
    &\bra{m}\mathbf{F}\cdot\mathbf{r}\ket{l}\bra{l}H_\mathrm{Z}\ket{l'}\bra{l'}S\ket{m'} \nonumber \\
    &\bra{m}\mathbf{F}\cdot\mathbf{r}\ket{l}\bra{l}S\ket{l'}\bra{l'}H_\mathrm{Z}\ket{m'}\,.
    \label{eq:couplings}
\end{align}
\YMN{Here the in-plane electric field $\mathbf{F}$ couples the HH ground state $\ket{m}=\ket{0,0,0,\pm 3/2}$ to HH excited states with the same $J_z$, $\ket{l}=\ket{1,0,n_z,\pm 3/2}$ ($\mathbf{F}\parallel\mathbf{x}$) or $\ket{l}=\ket{0,1,n_z,\pm 3/2}$ ($\mathbf{F}\parallel\mathbf{y}$). Note that the non-separable wave functions, Eq.~(\ref{eq:wavefun}), allow for a change of vertical quantum number $n_z$ even for an in-plane electric field. The $S$ term of $H_\mathrm{LK}$ as well as the in-plane magnetic field $\mathbf{B}_\parallel$ in $H_\mathrm{Z}$ couple HH $J_z=+\tfrac{3}{2}$ to LH $J_z=+\tfrac{1}{2}$ states, and HH $J_z=-\tfrac{3}{2}$ to LH $J_z=-\tfrac{1}{2}$ states; since heavy and light holes have different effective masses, hence different envelopes, $H_\mathrm{Z}$ and $H_\mathrm{LK}$ can couple different quantum numbers $n_x$, $n_y$, and $n_z$, so that most generally $\ket{l'}\equiv\ket{n_x,n_y,n_z',\pm 1/2}$. Finally, given the above chain of HH/LH couplings, the states $\ket{m}$ and $\ket{m'}$ must share the same $J_z$, so that Eqs.~(\ref{eq:couplings}) give rise to a $\propto FB_{\parallel}\sigma_z$ correction.} After some algebra, the effective drive Hamiltonian in the heavy-hole ground-state subspace actually reads
\begin{equation}
H_\mathrm{d}^\mathrm{eff}=(\mu_xF_xB_x+\mu_yF_yB_y)\cos(\omega_\mathrm{d}t)\sigma_z
\label{eq:Hdeff}
\end{equation}
with:
\begin{subequations}
\label{eq:mus}
\begin{align}
    \mu_x&=\frac{3\hbar^2e\gamma_3\kappa\mu_B}{m_0}\sum_{n_x,n_y,n_z,n_z'}\frac{I_1(n_z,0)}{\Delta^\mathrm{LH}_{n_x,n_y,n_z'}\Delta^\mathrm{HH}_{1,0,n_z}}\times \nonumber \\ \Big[&I_2(n_x,n_y,n_z',0,0,0)\Big(\sqrt{2}I_3(n_x,n_y,n_z',2,0,n_z)- \nonumber \\ &I_3(n_x,n_y,n_z',0,0,n_z)\Big) \nonumber \\
    +&I_2(n_x,n_y,n_z',1,0,n_z)I_3(n_x,n_y,n_z',1,0,0)\Big] \\
    \mu_y&=\frac{3\hbar^2e\gamma_3\kappa\mu_B}{m_0}\sum_{n_x,n_y,n_z,n_z'}\frac{I_1(n_z,0)}{\Delta^\mathrm{LH}_{n_x,n_y,n_z'}\Delta^\mathrm{HH}_{0,1,n_z}}\times \nonumber \\ \Big[&I_2(n_x,n_y,n_z',0,0,0)\Big(\sqrt{2}I_3(n_x,n_y,n_z',0,2,n_z)- \nonumber \\
    &I_3(n_x,n_y,n_z',0,0,n_z)\Big) \nonumber \\
    +&I_2(n_x,n_y,n_z',0,1,n_z)I_3(n_x,n_y,n_z',0,1,0)\Big]\,,
\end{align}
\end{subequations}
where we have introduced the splitting $\Delta^\mathrm{HH}_{n_x,n_y,n_z}$ between the HH ground state $\ket{0,0,0,J_z=\pm\tfrac{3}{2}}$ and the HH excited states $\ket{n_x,n_y,n_z,J_z=\pm\tfrac{3}{2}}$, the splitting $\Delta^\mathrm{LH}_{n_x,n_y,n_z}$ between the HH ground state and the LH states $\ket{n_x,n_y,n_z,J_z=\pm\tfrac{1}{2}}$, and the integrals:
\begin{subequations}
\label{eq:In}
\begin{align}
    &I_1(n_z',n_z)=\bra{Z_{n_z'}}\zeta(z)^{-1/2}\ket{Z_{n_z}} \\
    &I_2(n_x',n_y',n_z',n_x,n_y,n_z)=\langle\Psi_{n_x',n_y',n_z'}^l|\Psi_{n_x,n_y,n_z}^h\rangle\\
    &I_3(n_x',n_y',n_z',n_x,n_y,n_z)= \nonumber \\
    &\ \ \ \ \bra{\Psi_{n_x',n_y',n_z'}^l}\frac{\partial}{\partial z}\zeta(z)^{1/2}\ket{\Psi_{n_x,n_y,n_z}^h}\,.
\end{align}
\end{subequations}
Eq.~(\ref{eq:Hdeff}) accounts for the Rabi oscillations in a non-separable confinement potential with an in-plane magnetic field. \YMN{In particular, $\mu_x=\mu_y=0$ if $\zeta(z)=1$ because the $I_3$ integrals are all zero. Indeed, in such a separable potential, the in-plane electric field $\mathbf{F}$ and the Zeeman Hamiltonian $H_\mathrm{Z}$ can only couple the HH ground state $\ket{m}=\ket{0,0,0,\pm 3/2}$ to excited states $\ket{l}$ and $\ket{l'}$ with the same $n_z=0$, while $S$ can only couple $\ket{m'}\equiv\ket{m}$ to states $\ket{l'}$ with $n_z\ne0$. Hence there are no matching $\ket{l'}$ and no Rabi oscillations. The fact that $\mathbf{F}$ (and $H_\mathrm{Z}$) can couple non-separable wave functions with different $n_z$'s is therefore the key condition to achieve finite Rabi frequencies.}

The effective Hamiltonian of the ground-state heavy-hole doublet can therefore be approximated as: 
\begin{equation}
    {\cal H}\approx \frac{1}{2}\mu_B\boldsymbol{\sigma}\cdot \gt\mathbf{B}+\frac{1}{2}\mu_B V_\mathrm{ac}(\lambda_x B_x+\lambda_y B_y)\cos(\omega_\mathrm{d}t)\sigma_z\,,
\end{equation}
where $\lambda_x$, $\lambda_y$ relate $\mu_x$ and $\mu_y$ to the gate voltage drive $V_\mathrm{ac}$. We find that the corrections to Eqs.~(\ref{eq:gHH}) for the $\gt$-matrix resulting from the non-separability of the confinement potential are usually negligible. Strikingly, the last term can be cast as a time-dependent, off-diagonal $\gt$-matrix component leading to Eq.~(\ref{eq:gTMR}) of the main text for the Rabi frequency \cite{Venitucci18}.

The above conclusions have been supported by numerical calculations with separable and non-separable test potentials. These calculations confirm that there are no Rabi oscillations for in-plane magnetic fields and homogeneous AC electric fields unless the confinement potential is non-separable.

The scaling of the Rabi frequency with the size of the dot is pretty complex. The structure of Eq.~(\ref{eq:mus}) suggests that the Rabi frequency scales as $r_\parallel^2$ if the contribution from the $n_z=0$ term is dominant, owing to the $\Delta^\mathrm{HH}_{1,0,0}$ and $\Delta^\mathrm{HH}_{0,1,0}$ denominators (i.e., the Rabi oscillations are limited by the in-plane motion). On the other hand,  $\Delta^\mathrm{HH}_{1,0,n_z>0}$ and $\Delta^\mathrm{HH}_{0,1,n_z>0}$ are $\propto 1/L_\mathrm{W}^2$ due to the strong vertical confinement. Assuming $\zeta(z)\approx 1+\alpha z$ with $\alpha\ll 1/L_\mathrm{W}$, $I_1(n_z>0, 0)$ and the relevant $I_3$'s are then essentially dipole matrix elements that scale respectively as $\alpha L_\mathrm{W}$ and $\alpha^2 L_\mathrm{W}$. If $\Delta_\mathrm{LH}$ is ruled by strains, the $n_z>0$ terms hence make a $\propto \alpha^3 L_\mathrm{W}^4$ contribution to the sum-over-states. In the non-separable \YMN{confinement} potential of the device of the main text, we find a weak dependence of $f_\mathrm{R}$ on $r_\parallel$ but a quasi $\propto\ell_z^4$ behavior (see Fig.~\ref{fig:correlations}a), which suggests (as expected) that the Rabi frequency is more limited by the vertical than by the in-plane motion.

\section{$g$-TMR in a non-homogeneous AC electric field}
\label{app:gTMR}

In this appendix, we discuss \YMN{Rabi oscillations in a non-homogeneous AC electric field}.

\YMN{First of all, the non-separability of the AC drive potential can also contribute to $\gt_{zx}^\prime$ and $\gt_{zy}^\prime$, as does the non-separability of the confinement potential. This is best evidenced by Eqs.~(\ref{eq:couplings}). As discussed in Appendix \ref{app:theory}, a homogeneous AC electric field $\mathbf{F}$ can not couple states with different $n_z$'s if the confinement potential and wave functions are separable ($\zeta(z)=1$), which prevents connections between the HH ground states $\ket{m}\equiv\ket{m'}$. If the homogeneous AC electric field is replaced by an arbitrary $V_\mathrm{ac}(\mathbf{r},t)=V_\mathrm{ac}D(\mathbf{r})\cos\omega_\mathrm{d}t$, then $-e\mathbf{F}\cdot\mathbf{r}\to V_\mathrm{ac}D(\mathbf{r})$ \footnote{\YMN{If the electrostatics is linear (as is the case here since there are no hole gases around), $D(\mathbf{r})$ is nothing else that the potential created by a $1$\,V pulse on the driving gate(s) with all other gates grounded (e.g., $V_\mathrm{L}=-V_\mathrm{R}=0.5$\,V for opposite L/R drives). As $V_\mathrm{ac}(\mathbf{r},t)=V_\mathrm{ac}D(\mathbf{r})\cos\omega_\mathrm{d}t$ scales homogeneously with the drive amplitude $V_\mathrm{ac}$, there is no ``power onset'' for the non-separability of the AC potential, at least for Rabi frequencies to first-order in $V_\mathrm{ac}$.}}; if non-separable, $D(\mathbf{r})$ can generally couple the HH ground-state $\ket{m}=\ket{0,0,0,\pm 3/2}$ to HH excited states $\ket{l}=\ket{n_x,n_y,n_z,\pm 3/2}$ with $n_z\ne 0$ even if $\zeta(z)=1$, which restores the connections between $\ket{m}$ and $\ket{m'}$ in Eqs.~(\ref{eq:couplings}). As discussed in the main text, the NS mechanism however appears dominated in the present device by the non-separability of the confinement potential.}

Next, the principal $\gt$-factors $\gt_x$, $\gt_y$ and $\gt_z$ of a circular dot cannot be modulated (to first-order) by a homogeneous in-plane AC electric field (even if the static confinement potential is anharmonic) \cite{Venitucci18}. This follows from symmetries (see Tables \ref{tab:symmetries1} and \ref{tab:symmetries2}) and is consistent with the shape of the $\gt^\prime$ matrix for opposite L/R drives, whose diagonal is indeed zero. The principal $\gt$-factors can however be modulated by a non-homogeneous AC electric field that breaks parity, as is the case when the dot is driven by the L gate only.

\begin{table}
\begin{tabular}{|c|c|c|} \hline
 & $\sigma_{yz}$ & $\sigma_{xz}$ \\ \hline
$\mathbf{E}_\mathrm{ac}$ even &  
$\begin{pmatrix}
\bullet & 0 & 0 \\
0 & \bullet & \bullet \\
0 & \bullet & \bullet
\end{pmatrix}$ &
$\begin{pmatrix}
\bullet & 0 & \bullet \\
0 & \bullet & 0 \\
\bullet & 0 & \bullet
\end{pmatrix}$ \\ \hline
$\mathbf{E}_\mathrm{ac}$ odd & 
$\begin{pmatrix}
0 & \bullet & \bullet \\
\bullet & 0  & 0 \\
\bullet & 0 & 0
\end{pmatrix}$ & 
$\begin{pmatrix}
0 & \bullet & 0 \\
\bullet & 0 & \bullet \\
0 & \bullet & 0
\end{pmatrix}$  \\ \hline
Other & 
$\begin{pmatrix}
\bullet & \bullet & \bullet \\
\bullet & \bullet & \bullet \\
\bullet & \bullet & \bullet
\end{pmatrix}$ & 
$\begin{pmatrix}
\bullet & \bullet & \bullet \\
\bullet & \bullet & \bullet \\
\bullet & \bullet & \bullet
\end{pmatrix}$ \\
\hline
\end{tabular}
\caption{Constraints on the shape of $\gt^\prime$ set by the mirror planes  $\sigma_{yz}$ and $\sigma_{xz}$ of the device of Fig.~\ref{fig:device}, depending whether the AC electric field $\mathbf{E}_\mathrm{ac}=-\boldsymbol{\nabla}V_\mathrm{ac}$ is even [$\mathbf{E}_\mathrm{ac}(\sigma_{\alpha\beta}(\mathbf{r}))=\sigma_{\alpha\beta}(\mathbf{E}_\mathrm{ac}(\mathbf{r}))$], odd [$\mathbf{E}_\mathrm{ac}(\sigma_{\alpha\beta}(\mathbf{r}))=-\sigma_{\alpha\beta}(\mathbf{E}_\mathrm{ac}(\mathbf{r}))$], or does not show any relevant parity under that mirror transformation. The black dots are the non-zero matrix elements \cite{Venitucci18}.}
\label{tab:symmetries1}
\end{table}

\begin{table}
\begin{tabular}{|c|c|c|c|} \hline
\multirow{2}{*}{Drive} & Parity of $\mathbf{E}_\mathrm{ac}$ & Parity of $\mathbf{E}_\mathrm{ac}$ & \multirow{2}{*}{$\gt^\prime$} \\
 & wrt $\sigma_{yz}$ & wrt $\sigma_{xz}$ & \\
\hline
Homog. $\mathbf{E}_\mathrm{ac}\parallel\mathbf{x}$ & Odd & Even & $\begin{pmatrix}
0 & 0 & \bullet \\
0 & 0  & 0 \\
\bullet & 0 & 0
\end{pmatrix}$ \\ \hline
Opposite L/R & Odd & Even & $\begin{pmatrix}
0 & 0 & \bullet \\
0 & 0  & 0 \\
\bullet & 0 & 0
\end{pmatrix}$ \\ \hline
L & None & Even & $\begin{pmatrix}
\bullet & 0 & \bullet \\
0 & \bullet & 0 \\
\bullet & 0 & \bullet
\end{pmatrix}$ \\ \hline
C & Even & Even & $\begin{pmatrix}
\bullet & 0 & 0 \\
0 & \bullet & 0 \\
0 & 0 & \bullet
\end{pmatrix}$ \\
\hline
\end{tabular}
\caption{Shape of $\gt^\prime$ imposed by symmetries for the different drives considered in this work: homogeneous AC electric field $\mathbf{E}_\mathrm{ac}\parallel\mathbf{x}$, opposite drives on the L and R gates, drive on the L gate only, and on the C gate only (Appendix \ref{app:central}). The second and third columns are the parities of $\mathbf{E}_\mathrm{ac}$ with respect to the $\sigma_{yz}$ and $\sigma_{xz}$ mirrors. The last column is the shape of the $\gt^\prime$ constructed from the intersection of the relevant patterns of Table \ref{tab:symmetries1}. The diagonal elements describe conventional $g$-TMR; $\gt_{zx}^\prime$ describes the NS mechanism and $\gt_{xz}^\prime$ the cubic Rashba SOI.}
\label{tab:symmetries2}
\end{table}

\YMN{Given the structure of Eqs.~(\ref{eq:gHH}), this conventional $g$-TMR results from the interplay between the drive, the $R$ term of the LK Hamiltonian, and $H_\mathrm{Z}$ (or $S$ for the $\propto\eta_h$ contributions from the orbital motion in the magnetic field).} Let us write down the equations for the derivatives $\gt_x^\prime=\partial \gt_x/\partial V_\mathrm{L}$ and $\gt_y^\prime=\partial \gt_y/\partial V_\mathrm{L}$. First, the derivative of the ground-state heavy-hole wave function is:
\begin{equation}
\ket{\Psi_0^{h\prime}}=\frac{\partial\ket{\Psi_0^h}}{\partial V_\mathrm{L}}=\sum_{n>0}\frac{\bra{\Psi_n^h}D\ket{\Psi_0^h}}{E_0^h-E_n^h}\ket{\Psi_n^h}\,,
\label{eq:dPsidV}
\end{equation}
where $\Psi_n^h$ and $E_n^h$ are the heavy-hole wave functions and energies, and $D(\mathbf{r})=\partial V(\mathbf{r})/\partial V_\mathrm{L}$ is the derivative of the total potential $V(\mathbf{r})$ with respect to $V_\mathrm{L}$. The latter may be expanded around the origin (the center of the dot) as:
\begin{equation}
D(\mathbf{r})= D_0+\mathbf{D}_1\cdot\mathbf{r}+\frac{1}{2}\mathbf{r}\cdot\mathsf{D}_2\mathbf{r}+{\cal O}(\mathbf{r}^2)\,,
\label{eq:expansion}
\end{equation}
where $\mathbf{D}_1$ is the AC electric field (per unit $\delta V_\mathrm{L}$) and $\mathsf{D}_2$ is the Hessian matrix of $D(\mathbf{r})$ at the origin. As discussed above, a homogeneous electric field cannot modulate the principal $\gt$-factors of a circular dot, so that $D_0$ and $\mathbf{D}_1$ [and actually all odd powers in Eq.~(\ref{eq:expansion})] do not contribute to $\gt_x^\prime$ and $\gt_y^\prime$ \footnote{It is worthwhile, for the sake of clarity, to point out an important difference between the present Ge devices and the Si devices of Ref.~\cite{Venitucci18}: in the former, the Rabi frequency remains non-zero when the wave function is symmetric, whereas in the latter, the Rabi frequency goes through an exact zero at some bias where the wave functions are ```quasi'' (yet not even exactly) symmetric -- although the drive is non-homogeneous in both cases. However, a close inspection of Fig. 13 of Ref.~\cite{Venitucci18} reveals that the relevant component of $\mathsf{D}_2$ (namely, $\partial^2 V_\mathrm{ac}/\partial y^2$) is zero for that particular overlapping gate layout. In other words, a positive and negative drive respectively squeeze and enlarge the dot in the Ge devices, but both squeeze the dot symetrically in the Si devices near the zero of $f_\mathrm{R}$.}. If the dot moves essentially in-plane, the relevant matrix elements of $\mathsf{D}_2$ shall scale as $r_\parallel^2$, while the energy denominators in Eq.~(\ref{eq:dPsidV}) shall scale as $1/r_\parallel^2$. Therefore, $\ket{\Psi_0^{h\prime}}$ is expected to scale as $r_\parallel^4$, so that $\partial\langle p_i^2\rangle/\partial V_\mathrm{L}=\bra{\Psi_0^h}p_i^2\ket{\Psi_0^{h\prime}}+\mathrm{c.c.}$ scales as $r_\parallel^2$.

Hence $\gt_x^\prime$ and $\gt_y^\prime$ also scale as $r_\parallel^2$, and so does the Rabi frequency for an in-plane magnetic field [Eq.~(\ref{eq:fRg})],
\begin{equation}
f_\mathrm{R}=\frac{\mu_BBV_\mathrm{ac}}{2h}|\gt_x^\prime\YMN{+}\gt_y^\prime||b_xb_y|\,.
\label{eq:fRTMRp}
\end{equation}
This expression is maximal when $b_x=b_y=\pm 1/\sqrt{2}$ \footnote{The Rabi frequency is also maximum (though slightly different) when the in-plane magnetic field lies $45^\circ$ from the axes of the side gates if the latter are oriented along $\{110\}$.}:
\begin{equation}
f_\mathrm{R}(\mathbf{B}\parallel\pm\mathbf{x}\pm\mathbf{y})=\frac{\mu_BBV_\mathrm{ac}}{4h}|\gt_x^\prime\YMN{+}\gt_y^\prime|\,.
\label{eq:fRTMRp2}
\end{equation}
The conventional $g$-TMR is however optimal (at constant magnetic field) when $\mathbf{B}$ lies in the $xz$ plane, where:
\begin{equation}
f_\mathrm{R}=\frac{\mu_BBV_\mathrm{ac}}{2h}|\gt_z\gt_x^\prime-\gt_z^\prime\gt_x|\frac{|b_xb_z|}{\sqrt{\gt_x^2b_x^2+\gt_z^2b_z^2}}\,.
\end{equation}
The above Rabi frequency peaks when the magnetic field makes an angle $\theta^\pm=\pi/2\pm\arctan\sqrt{\gt_x/\gt_z}$ with the $z$ axis (see Fig.~\ref{fig:VgmapTMR}a) \cite{Venitucci19,Michal21}, where, assuming $\gt_z\gg\gt_x$,
\begin{equation}
f_\mathrm{R}^\mathrm{max}\approx\frac{\mu_BBV_\mathrm{ac}}{2h}|\gt_x^\prime|\,.
\end{equation}
\YMN{This expression is slightly larger than Eq.~(\ref{eq:fRTMRp2}) because $|\gt_x^\prime|\gtrsim|\gt_y^\prime|$ owing to the strong contribution of the $\propto\eta_h$ terms in the derivatives of $\gt_{x,y}$ [Eqs. (\ref{eq:gHH})]. This underlines the role of the orbital motion in the magnetic vector potential}. At constant Larmor frequency, the fastest Rabi oscillations are, nonetheless, achieved with an in-plane magnetic field in the device of Fig.~\ref{fig:device} owing to the large $\gt_z/\YMN{|}\gt_{x,y}\YMN{|}$ ratio ($\gt^*=\gt_\parallel$ at $\theta=\pi/2$ and $\gt^*=\sqrt{\gt_\parallel\gt_z}$ at $\theta=\theta^\pm$).

\section{Manipulation with the central gate}
\label{app:central}

\begin{figure}
    \centering
    \includegraphics[width=1\columnwidth]{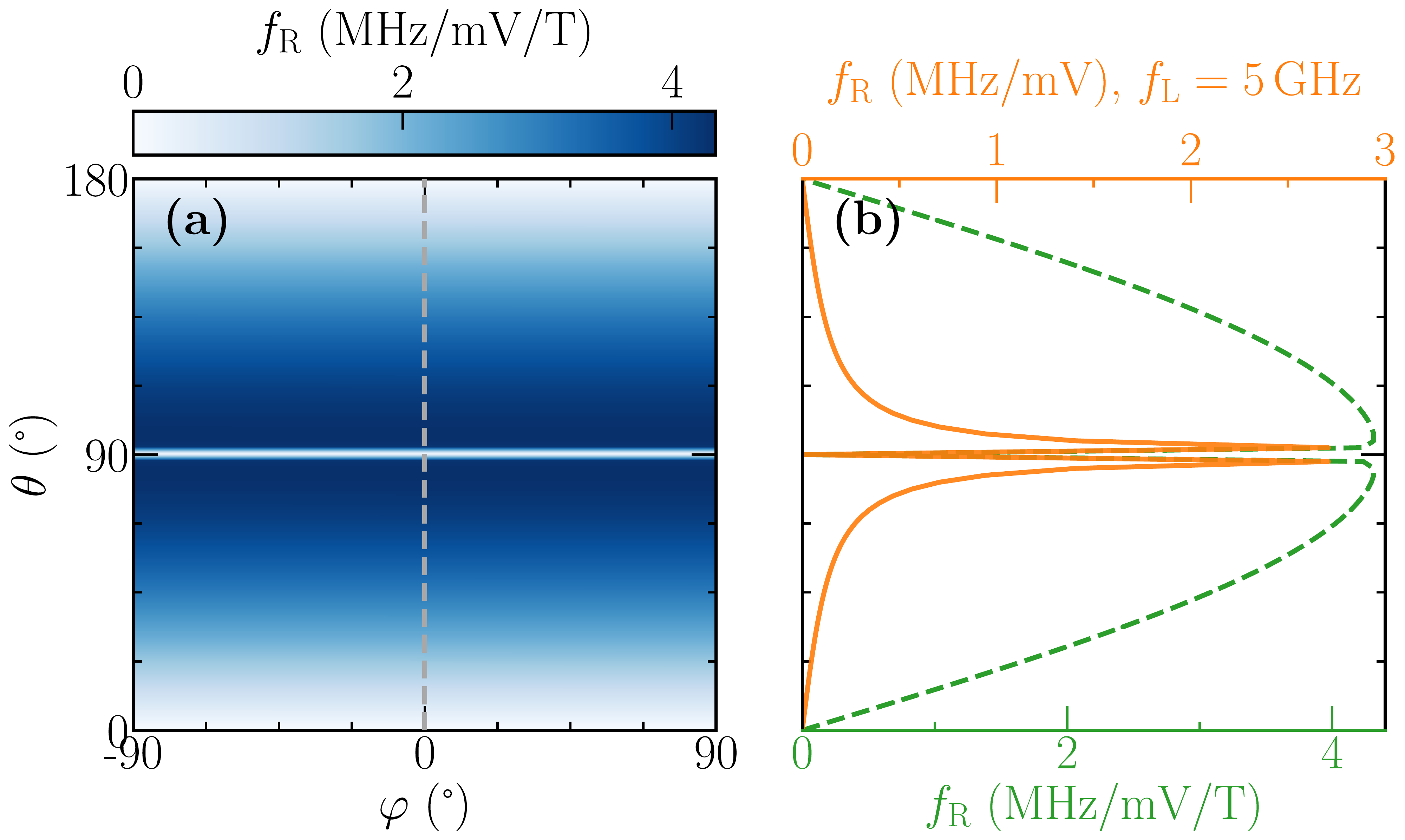}
    \caption{(a) Map of the Rabi frequency $f_\mathrm{R}$ as a function of the magnetic field angles $\theta$ and $\varphi$ defined in Fig.~\ref{fig:device}, for a drive $\delta V_\mathrm{C}=V_\mathrm{ac}\cos\omega_\mathrm{L}t$ on the C gate ($V_\mathrm{C}=-40$\,mV, $V_\mathrm{bg}=0$\,V, $V_\mathrm{ac}=1$\,mV and $B=1$\,T). (b) Cut along the dashed gray line in (a), at constant magnetic field $B=1$\,T (green), and at constant Larmor frequency $f_\mathrm{L}=\omega_\mathrm{L}/(2\pi)=5$\,GHz (orange).}
    \label{fig:central}
\end{figure}

In this appendix, we briefly address manipulation with the central gate C. The Rabi frequency is plotted as a function of the magnetic field angles $\theta$ and $\varphi$ in Fig.~\ref{fig:central}, for a drive $\delta V_\mathrm{C}=V_\mathrm{ac}\cos\omega_\mathrm{L}t$ on the C gate. It is significantly smaller than for a drive with the L gate (Fig.~\ref{fig:VgmapTMR}), and is, moreover, zero for in-plane magnetic fields.

The drive on the C gate does not break any symmetry and can, therefore, only modulate the principal $g$-factors $\gt_x$, $\gt_y$ and $\gt_z$ (owing to the ``breathing'' of the dot ruled by $\mathsf{D}_2$ in Eq.~(\ref{eq:dPsidV}), see also Table \ref{tab:symmetries2}). Defining $\gt_\parallel=\gt_x=\YMN{-}\gt_y$, and $\gt_\parallel^\prime=\partial\gt_x/\partial V_\mathrm{C}=\YMN{-}\partial\gt_y/\partial V_\mathrm{C}$, the Rabi frequency [Eq.~(\ref{eq:fRg})] reads for an arbitrary magnetic field orientation:
\begin{equation}
f_\mathrm{R}=\frac{\mu_BBV_\mathrm{ac}}{2h}|\gt_z\gt_\parallel^\prime-\gt_\parallel\gt_z^\prime|\frac{|b_zb_\parallel|}{\sqrt{\gt_\parallel^2b_\parallel^2+g_z^2b_z^2}}\,,
\end{equation}
where $b_\parallel=\sqrt{b_x^2+b_y^2}$. It is, therefore, independent on the polar angle $\varphi$, and is expected to be zero when the magnetic field lies in-plane or along $z$. Slight symmetry breakings  (if the dot is not perfectly centered below the C gate for example) might give rise to finite, yet slow Rabi oscillations for in-plane magnetic fields. As in Appendix \ref{app:gTMR}, the Rabi frequency is maximum when $\theta=\theta^\pm=\pi/2\pm\arctan\sqrt{\gt_\parallel/\gt_z}$, and reaches (assuming $\gt_z\gg\gt_\parallel$):
\begin{equation}
f_\mathrm{R}^\mathrm{max}\approx\frac{\mu_BBV_\mathrm{ac}}{2h}|\gt_\parallel^\prime|\,.
\end{equation}

\bibliography{biblio}

\end{document}